\documentclass{article}

\usepackage{arxiv}

\usepackage[utf8]{inputenc} % allow utf-8 input
\usepackage[T1]{fontenc}    % use 8-bit T1 fonts
\usepackage{hyperref}       % hyperlinks
\usepackage{url}            % simple URL typesetting
\usepackage{booktabs}       % professional-quality tables
\usepackage{amsfonts}       % blackboard math symbols
\usepackage{nicefrac}       % compact symbols for 1/2, etc.
\usepackage{microtype}      % microtypography
\usepackage{lipsum}		% Can be removed after putting your text content
\usepackage{graphicx}
\usepackage[style=ieee, citestyle=numeric-comp, backend=biber, dashed=false]{biblatex}
\usepackage{doi}
\usepackage{bm}
\usepackage{float}
\usepackage{wrapfig}
\usepackage{makecell}
\usepackage{bm}
\usepackage{rotating}
\usepackage[flushleft]{threeparttable}
\usepackage{multirow}
\usepackage{algorithm}
\usepackage{algpseudocode}

\usepackage{amsmath,amssymb}

\DeclareMathOperator*{\argmin}{argmin}

\title{Automatic differentiation accelerated shape optimization approaches to photonic inverse design on rectilinear simulation grids}
\author{ \href{https://orcid.org/0000-0003-1260-412X}{\includegraphics[scale=0.06]{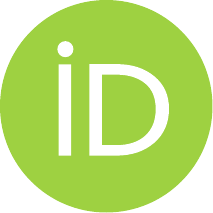}\hspace{1mm}Sean~Hooten}
		\\
	Hewlett Packard Labs \\
	Hewlett Packard Enterprise \\
	Milpitas, CA 95035, USA \\
	\texttt{sean.hooten@hpe.com} \\
 	\And
        Peng Sun \\
	NVIDIA Corporation \\
	2788 San Tomas Expy.\\
    Santa Clara, CA 95051, USA \\
	\texttt{pesun@nvidia.com} \\
	\And
    Liron Gantz \\
	NVIDIA Corporation \\
	Hakidma 26 \\
	Yokneam, Israel \\
	\texttt{lironga@nvidia.com} \\
	\And
    Marco Fiorentino \\
	Hewlett Packard Labs\\
	Hewlett Packard Enterprise \\
	Milpitas, CA 95035, USA \\
	\texttt{marco.fiorentino@hpe.com} \\
        \And
    Raymond G. Beausoleil \\
	Hewlett Packard Labs\\
	Hewlett Packard Enterprise \\
	Milpitas, CA 95035, USA \\
	\texttt{ray.beausoleil@hpe.com} \\
	\And
	\href{https://orcid.org/0000-0002-7301-8610}{\includegraphics[scale=0.06]{orcid.pdf}\hspace{1mm}Thomas~Van Vaerenbergh} \\
	Hewlett Packard Labs\\
	HPE Belgium\\
	B-1831 Diegem, Belgium \\
	\texttt{thomas.van-vaerenbergh@hpe.com}
}

\renewcommand{\shorttitle}{Accelerated shape optimization via automatic differentiation}

\hypersetup{
pdftitle={Accelerated shape optimization approaches to photonic inverse design on rectilinear simulation grids via automatic differentiation},
pdfsubject={inverse design, electromagnetics, optimization, automatic differentiation},
pdfauthor={Sean~Hooten},
pdfkeywords={inverse design, adjoint variables method, shape optimization, structural optimization, machine learning, automatic differentiation, electromagnetism, integrated photonics, optics, finite-difference time-domain (FDTD)},
}

\begin{document}
\maketitle

\begin{abstract}
    Shape optimization approaches to inverse design offer low-dimensional, physically-guided parameterizations of structures by representing them as combinations of shape primitives. However, on discretized rectilinear simulation grids, computing the gradient of a user objective via the adjoint variables method requires a sum reduction of the forward/adjoint field solutions and the Jacobian of the simulation material distribution with respect to the structural shape parameters. These shape parameters often perturb large or global parts of the simulation grid resulting in many non-zero Jacobian entries, which are typically computed by finite-difference in practice. Consequently, the gradient calculation can be non-trivial. In this work we propose to accelerate the gradient calculation by invoking automatic differentiation (AutoDiff) in instantiations of structural material distributions. In doing so, we develop extensible differentiable mappings from shape parameters to shape primitives and differentiable effective logic operations (denoted AutoDiffGeo). These AutoDiffGeo definitions may introduce some additional discretization error into the field solutions because they relax notions of sub-pixel smoothing along shape boundaries. However, we show that some mappings (e.g. simple cuboids) can achieve zero error with respect to volumetric averaging strategies. We demonstrate AutoDiff enhanced shape optimization using three integrated photonic examples: a multi-etch blazed grating coupler, a non-adiabatic waveguide transition taper, and a polarization-splitting grating coupler. We find accelerations of the gradient calculation by AutoDiff relative to finite-difference often exceed $50\times$, resulting in total wall time accelerations of $4\times$ or more on the same hardware with little or no compromise to final device performance. Our code is available open source at \texttt{https://github.com/smhooten/emopt}.
\end{abstract}

% keywords can be removed

\section{Introduction}
\label{sec:introduction}
%In shape optimization approaches to inverse design via the adjoint method, physical structures are represented as combinations of elementary shape primitives, such as polygons or cylinders. 
For several decades the adjoint variables method has been celebrated in the photonics community for enabling high-dimensional, computationally-efficient optimizations of electromagnetic structures \cite{ bendsoe_material_1999, cao_adjoint_2003, bendsoe_topology_2004, jensen_topology_2011,  Miller:EECS-2012-115, peter_numerical_2010, kao_maximizing_2005,  wang_robust_2011, elesin_design_2012, lu_objective-first_2012, shen_integrated-nanophotonics_2015, molesky_inverse_2018, michaels_leveraging_2018, michaels_inverse_2018, hughes_forward-mode_2019, andrade_inverse_2019, michaels_hierarchical_2020,     piggott_inverse-designed_2020, zhang_experimental_2022, minkov_inverse_2020, sun_adjoint_2023, ahn_augmenting_2021, zhang_enabling_2021, hammond_high-performance_2022,  christiansen_inverse_2020, vercruysse_inverse-designed_2021, hammond_multi-layer_2022, hammond_photonic_2021, christiansen_topological_2019, schubert_inverse_2022, cecil_advances_2023, lou_inverse_2023,  wang_adjoint-based_2018, veronis_method_2004, piggott_inverse_2015-1, vercruysse_analytical_2019, su_nanophotonic_2019, sapra_inverse_2019, zhou_minimum_2015, lu_nanophotonic_2013,  hooten_adjoint_2020, hammond_phase-injected_2023, chen_design_2020, gershnabel_reparameterization_2023}. Various adjoint method approaches differ in their representation of device geometry on rectilinear simulation grids used in popular algorithms such as finite-difference frequency-domain (FDFD) and finite-difference time-domain (FDTD). These representations can be broadly classified in ``topology'', ``level-set'', or ``shape'' optimization categories. This paper proposes a software acceleration for the latter shape optimization category by invoking automatic differentiation (AutoDiff) \cite{rumelhart_learning_1986, paszke2019pytorch} in instantiations of structural material distributions. 

\begin{figure}
    \centering
    \includegraphics[width=0.75\textwidth]{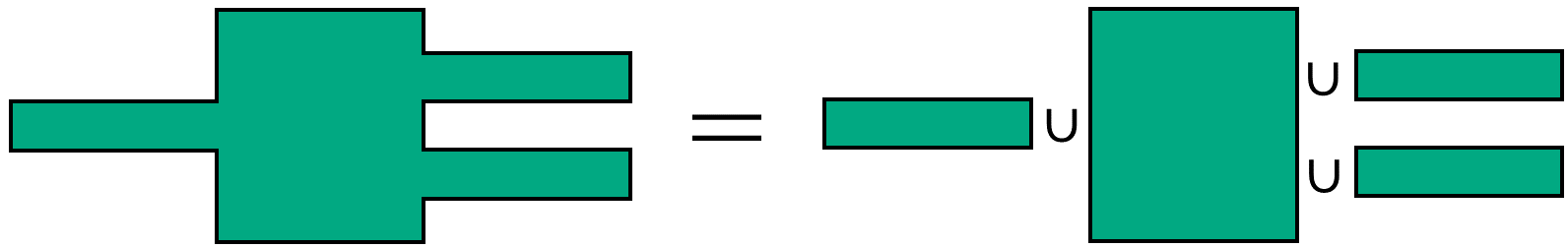}
    \caption{2D multi-mode-interferometer-like (MMI-like) structure consists of the union of 4 rectangles.}
    \label{fig:ASO_example}
\end{figure}

In shape optimization approaches to inverse design, photonic structures are represented by logical combinations of elementary shape primitives, such as cuboids or cylinders.  Shapes have interiors and exteriors, and strictly speaking, we define shapes here as closed sets in $\mathbb{R}^d$ (where $d=2,3$ for 2D, 3D) with parametrically-defined boundaries. From the perspective of electromagnetic simulation, every point sampled from the interior of a shape has associated electromagnetic material properties (such as the linear permittivity and permeability, $\varepsilon$ and $\mu$). Shape representations of photonic structures are advantageous from an engineer's point-of-view, because complicated photonic structures can be easily assembled from logically-defined combinations of individual shape primitives. Shape representations are also commonly used for integrated circuit layout definitions (e.g., in GDSII file format). A simple example of a shape optimization compatible multi-mode-interferometer-like (MMI-like) structure is shown in Fig.\,\ref{fig:ASO_example}, which is assembled via the logical union of four simpler elementary rectangle primitives.
Owing to their limited degrees-of-freedom, shape representations of photonic structures provide the additional advantages of (i) physical interpretability \cite{michaels_hierarchical_2020}, (ii) solutions that automatically satisfy sub-pixel smoothing  \cite{farjadpour_improving_2006,oskooi_accurate_2009,michaels_leveraging_2018} for minimization of discretization error, (iii) the ability to easily define geometrical constraints, e.g. lithographic minimum feature size \cite{michaels_hierarchical_2020, chen_design_2020, gershnabel_reparameterization_2023}, (iv) predictable fabrication biases, and (v) dimensionality-reduction techniques for adjoint-inspired design-of-experiments \cite{vaerenbergh_wafer-level_2021, vaerenbergh_wafer-level_2022}. 

In this paper we concern ourselves with shape optimization on rectilinear discretized simulation grids, commonly used in finite-difference time domain (FDTD) methods. This situation presents a problem for the gradient-based inverse design algorithm, which results from the lack of a closed-form expression for the material properties along each element of the simulation grid. %In other words, currently the only way to define the material property of any given simulation grid element is to exhaustively check whether it is inside, outside, or on the boundary of each user-defined shape. This amounts to a sequence of logical expressions (``if inside shape A, but not shape B: grid element takes on permittivity value $\alpha$, ...''). 
In particular, the material value assigned to any given grid element is chosen by volumetrically averaging the material values of the various shapes that occupy that grid element in order to minimize discretization error and to assuage discontinuities in material values of nearby grid cells, called boundary smoothing or sub-pixel smoothing \cite{farjadpour_improving_2006, oskooi_accurate_2009, oskooi_meep_2010, michaels_leveraging_2018}. This amounts to performing numerical quadrature over the intersection of shapes with each grid cell. Sub-pixel smoothing leads to physically well-behaved and well-defined material Jacobians: $\partial \varepsilon_j/\partial v_i$ and $\partial \mu_j/\partial v_i$, where $\varepsilon_j$ and $\mu_j$ are the material properties of grid element $j$ and $v_i$ is shape parameter $i$. However, all practical implementations known to the authors to generate the material Jacobians of exact shape representations \cite{anstmichaels_emopt_2023, lumerical, flexcompute_python-driven_nodate, noauthor_nanocompmeep_2023} are by finite-difference, where material derivatives are computed manually under shape parameter perturbations  (e.g., $\partial \varepsilon_j/\partial v_i \approx (\varepsilon_j(v+\delta v_i) - \varepsilon_j(v))/{\delta v_i}$). As we will show, computing the adjoint method gradient in general requires that the material Jacobians are known across all parameters ($i=1,...,n$) and all simulation grid elements within some design region ($j=1,...,m$). Consequently, in the worst case, $\varepsilon$ and $\mu$ must be recomputed $n$ times over $m$ grid elements -- a nontrivial calculation especially in large, 3D simulation domains.

This paper proposes to accelerate the calculation of the adjoint method gradient (stated in Sec.\,\ref{sec:background} and modified for AutoDiff in Sec.\,\ref{sec:autodiff}) by representing elementary shape primitives and Boolean logic operations with closed-form, differentiable expressions, which we call AutoDiffGeo (detailed in Sec.\,\ref{sec:autodiff} and Sec.\,\ref{sec:diffgeo}). The key to making this possible %is to relax the notion of exact sub-pixel smoothing over a shape with nominally discontinuous boundary representations, and instead to 
is to directly define material properties as scalar fields with smooth transitions along parametric boundaries.
These smooth boundary representations can lead to a deviation from exact sub-pixel smoothing of material properties (i.e., non-physical `grayscale' $\varepsilon$ and $\mu$, commonly seen in topology optimization before strict binarization of material values is enforced \cite{lebbe_robust_2019, piggott_inverse-designed_2020, hammond_photonic_2021, zhang_experimental_2022}). However, in Sec.\,\ref{sec:subpixelapprox} we demonstrate that many common shapes used in typical design flows (e.g. cuboids and planarity transformations) can achieve zero error with respect to common volumetric-averaging sub-pixel smoothing strategies, and we address the arbitrary shape cases with multiple proposals in the remainder of Sec.\,\ref{sec:subpixelsmoothing}. Finally, in Sec.\,\ref{sec:experiments} we perform several inverse design optimizations on practical use cases leveraging AutoDiffGeo shape representations, with comparison to standard finite-difference techniques. We find accelerations of the adjoint method gradient calculation exceeding >$50\times$, resulting in total wall time acceleration >$4\times$ in our test cases.

\paragraph{Related work}
Photonic inverse design has and continues to be a subject of extremely active research. Of the adjoint variables method based approaches, the vast majority of recent works implement density-based topology optimization representations of photonic structures \cite{jensen_topology_2011, piggott_inverse-designed_2020, zhang_experimental_2022, ahn_augmenting_2021, hammond_photonic_2021, hammond_multi-layer_2022, hammond_high-performance_2022, hammond_phase-injected_2023, christiansen_inverse_2020, vercruysse_inverse-designed_2021, schubert_inverse_2022, lou_inverse_2023, wang_adjoint-based_2018, piggott_inverse_2015-1, sapra_inverse_2019, hughes_adjoint_2018, zhou_minimum_2015, shen_integrated-nanophotonics_2015, jenkins_general-purpose_2023, shang_inverse-designed_2023}, with many recent theoretical advances successfully implementing fabrication compatibility \cite{hammond_photonic_2021, hammond_multi-layer_2022, hammond_high-performance_2022, zhang_enabling_2021, schubert_inverse_2022, jenkins_general-purpose_2023, shang_inverse-designed_2023}. Level-set \cite{kao_maximizing_2005, vercruysse_analytical_2019, lebbe_robust_2019} and shape optimization \cite{Miller:EECS-2012-115, lalau-keraly_adjoint_2013, michaels_inverse_2018, michaels_hierarchical_2020, hooten_adjoint_2020, chen_design_2020, vaerenbergh_wafer-level_2022, sun_adjoint_2023, gershnabel_reparameterization_2023, minkov_inverse_2020} approaches have seen less attention, but we note that most modern commercial and open source FDFD/FDTD simulation and optimization tools support shape optimization, such as Ansys Lumerical \cite{lumerical}, FlexCompute Tidy3D \cite{flexcompute_python-driven_nodate}, MEEP \cite{oskooi_meep_2010, noauthor_nanocompmeep_2023}, and EMopt \cite{michaels_leveraging_2018, anstmichaels_emopt_2023}. The AutoDiffGeo representations proposed here can be viewed as a mixture of level-set and shape representations, but with particular attention paid to smoothing boundary errors of shapes defined on rectilinear simulation grids. We note that similar differentiable representations of some shapes have been proposed in Refs.\,\cite{chen_design_2020, gershnabel_reparameterization_2023, zhelyeznyakov_large_2023} and in MEEP's documentation \cite{oskooi_meep_2010, noauthor_nanocompmeep_2023}, but to our knowledge, no other works have proposed (i) a large, extensible library of differentiable shape objects, (ii) differentiable effective logic operations such as intersection and union applied to those objects, and (iii) methods to improve or perfect sub-pixel smoothing of those shape representations as defined on rectilinear simulation grids (in particular, we reveal that some functionally-defined shapes can obtain zero error with respect to volumetric averaging, making these shape representations viable with otherwise large grid discretization sizes). Furthermore, we note that all shape representations discussed here can be extended with ``reparameterization'' \cite{chen_design_2020, gershnabel_reparameterization_2023} in order to obtain objects that automatically satisfy fabrication constraints. 

Several recent works have proposed accelerations to photonic inverse design. This includes automatic differentiation approaches (that are fundamentally different from the approach in this work) for circuit design \cite{gu_adept_2022, laporte_highly_2019}, photonic crystals \cite{minkov_inverse_2020}, nonlinear devices \cite{rozenberg_inverse_2022, hamerly_design_2022}, and multi-objective design tasks \cite{hughes_forward-mode_2019, noauthor_ceviche_2023}, as well as improvements to the underlying electromagnetic simulator by hardware \cite{lumerical, noauthor_fdtd-z_2023, peng_gpuemopt_2023, minkov_hardware-accelerated_2023} and algorithmic \cite{trivedi_data-driven_2019, dasdemir_computational_2023} acceleration. %Furthermore, a forward-mode automatic differentiation approach has been suggested to accelerate multi-objective design tasks \cite{hughes_forward-mode_2019, noauthor_ceviche_2023}. 
The adjoint variables method itself can be viewed as an application of reverse-mode automatic differentiation for scalar objectives \cite{hughes_forward-mode_2019}, for which density-based topology optimization is the most straightforward special case (which we will show in Sec.\,\ref{sec:autodiff}). Thus, we clarify that the acceleration due to automatic differentiation suggested in this work is through the use of AutoDiff-compatible smooth shape definitions. Conventionally, in shape optimization on rectilinear simulation grids, the adjoint variables method propagates sensitivities from the objective to the simulation grid, then finite-differences propagates those sensitivities to the structural shape parameters. We insert differentiable shape representations to accelerate the latter step (meanwhile providing exact numerical derivatives). Indeed, the shape representations in this paper can plug-and-play with any external photonic optimization tool with built-in forward and adjoint simulators.

We note that this work exclusively focuses on ``discretize before differentiate'' approaches to shape optimization on rectilinear simulation grids. This case presents the unique problem that the discretized material Jacobian must be computed in order to evaluate the gradient via the adjoint method. The analytic/continuous version of shape optimization \cite{Miller:EECS-2012-115, lalau-keraly_adjoint_2013} does not suffer from this problem (but requires a continuous field solution, which may not be available in general use cases). Furthermore, the acceleration suggested in this work is not necessarily relevant to finite-element methods (FEM), because there the continuous shape optimization case can be closely approximated through careful meshing.

While not directly relevant to this paper, we note that several recent works propose alternative or enhanced methods for photonic inverse design, namely machine learning and artificial intelligence approaches \cite{laporte_highly_2019, gostimirovic_improving_2023, gu_adept_2022, yeung_elucidating_2020, yeung_deepadjoint_2022, woldseth_use_2022, zhelyeznyakov_large_2023, rozenberg_inverse_2022, hooten_inverse_2021, yang_normalizing_2023, lu_physics-informed_2021, chen_high_2022}, and scalable Green's function approaches \cite{boutami_efficient_2019, boutami_efficient_2019-1, wang_massively_2022}. The shape representations proposed in this work are immediately applicable to several of the cited works, and are especially complementary to methods utilizing neural networks by virtue of their compatibility with automatic differentiation and backpropagation.

\paragraph{Our contributions}
\begin{itemize}
    \item A demonstration and practical discussion of how automatic differentiation can enhance shape optimization approaches to inverse design in FDTD.
    \item An extensible library of AutoDiff-compatible shape representations and effective differentiable logic operations (called AutoDiffGeo), which support non-binary material levels.
    \item An extended discussion and multiple proposals for how to minimize boundary errors with respect to exact sub-pixel smoothing on rectilinear simulation grids, including AutoDiffGeo definitions that automatically satisfy volumetric averaging.
    \item Multiple integrated photonic experiments demonstrating the utility of AutoDiff-enhanced shape optimization.
    \item A photonic topology optimization and AutoDiff-enhanced shape optimization toolkit are implemented open source in EMopt \cite{hooten_emopt_2023}.
\end{itemize}

\section{Background on photonic inverse design and the adjoint method}
\label{sec:background}
In this section we provide a brief mathematical description of inverse design in the context of common photonics applications. More general derivations can be found in Refs.\,\cite{cao_adjoint_2003, michaels_leveraging_2018, molesky_inverse_2018, bradley_pde-constrained_2019, Miller:EECS-2012-115}. However, we note that the geometry representations given in this paper are general and can be applied outside of the limited photonic applications treated here.

Suppose we are interested in designing a photonic device on a platform that consists only of linear diagonally-anisotropic materials, such as a common silicon-on-insulator (SOI) platform. Furthermore, we will only consider time-harmonic Maxwell's Equations with frequency, $\omega$. Let $\mathbf{v}\in\mathbb{R}^n$ represent an $n$-dimensional optimization variable, where each element $v_i\in\mathbf{v}$ corresponds to a geometrical degree-of-freedom in a photonic shape optimization problem, such as the length of a silicon rectangle. In shape optimization approaches to inverse design, we intend to optimize an electromagnetic objective with respect to $\mathbf{v}$, i.e. $\mathbf{v^*}=\argmin_\mathbf{v} f(\mathbf{E},\mathbf{H})$, where $f$ is a user-defined scalar objective function that is differentiable with respect to the electric and magnetic fields $\mathbf{E},\mathbf{H}$. For simplicity we have dropped explicit dependence of $f$ with respect to $\mathbf{v}$, $\mathbf{\varepsilon}$, and $\mu$, but we note that the results in this paper are easily extended to those cases.%Without loss of generality, we note that the electric and magnetic fields are typically discretized on a simulation grid with $m$ cells, i.e. each quantity can be written as an $m$-tuple: $\mathbf{E}=\{\mathbf{E}_1,\cdots,\mathbf{E}_m\}$, where each element of this tuple may

This optimization problem can be solved by gradient-descent, where the gradient is efficiently obtained using the adjoint variables method. Assuming discretized Maxwell's equations, with time-harmonic fields, linear isotropic materials (or diagonally anisotropic), and no dependence of current sources with $\mathbf{v}$, the derivative of $f$ w.r.t. each design variable $v_i\in \mathbf{v}$ can be expressed as \cite{michaels_leveraging_2018}:
\begin{align}
    \frac{\partial f}{\partial v_i}= 2 \omega \text{Im}\left(\sum_{j=1}^m \frac{\partial \varepsilon_{j}}{\partial v_i} {E}_{j} {E}_{j}^{\text{adj}}
    - \frac{\partial \mu_{j}}{\partial v_i} {H}_{j} {H}_{j}^\text{adj} \right),\quad i=1,\ldots,n
    \label{eq:1}
\end{align}
where $\omega$ is the frequency, $\varepsilon$ is the permittivity, $\mu$ is the permeability, and $\mathbf{E}^\text{adj}, \mathbf{H}^\text{adj}$ are the adjoint electric and magnetic fields (found by solving a system of equations that resemble Maxwell's Equations). Note that we sum-reduce over all spatial pixels/voxels in the simulation design region, as well as each polarization (usually defined at staggered grid locations in FDFD and FDTD), all lumped into index $j=1,...,m$. Immediately it can be observed that the derivatives of $f$ with respect to all $v_i$ variables can be computed with just two simulations: the ``forward'' simulation ($\mathbf{E},\mathbf{H}$) and the ``adjoint'' simulation ($\mathbf{E}^\text{adj}, \mathbf{H}^\text{adj}$). Consequently, a typical adjoint method based photonic inverse design protocol can be understood as a series of 7 (simplified) steps:
\begin{enumerate}
    \item Initialize $\mathbf{v}$.
    \item Compute the material distributions: $\mathbf{v}\mapsto\varepsilon, \mu$.
    \item Compute the forward simulation: $\mathbf{E},\mathbf{H}$.
    \item Compute the adjoint simulation: $\mathbf{E}^\textrm{adj},\mathbf{H}^\textrm{adj}$.
    \item Compute the gradient (Eq.\,\ref{eq:1}): $\nabla_\mathbf{v}f$.
    \item Update $\mathbf{v}$ using a gradient-descent-based protocol: $\mathbf{v}\leftarrow\mathbf{v}-\alpha\nabla_\mathbf{v}f$, for $\alpha>0$.
    \item Repeat steps (2)--(6) until a desired convergence criterion is satisfied, such as $\nabla_\mathbf{v}f\approx 0$.
\end{enumerate}

Note that this paper considers $\mathbf{E},\mathbf{H},\mathbf{E}^\text{adj}, \mathbf{H}^\text{adj}$ to be given quantities. In other words, finding improvements to the forward and adjoint simulators is out-of-scope. We refer the reader to algorithmic accelerations in Refs.\,\cite{trivedi_data-driven_2019, dasdemir_computational_2023} and to recent incorporations of GPU hardware acceleration \cite{lumerical, noauthor_fdtd-z_2023, peng_gpuemopt_2023, flexcompute_python-driven_nodate} in FDTD for these topics. This paper is only concerned with accelerating steps 2 and 5.

\section{Automatic differentiation based acceleration of adjoint variables method gradient}
\label{sec:autodiff}
In conventional shape optimization approaches to inverse design in FDTD, the quantities $\varepsilon$ and $\mu$ are computed by performing geometrical Boolean operations or exhaustive numerical quadrature between the rectilinear simulation grid elements and elementary shape primitives (e.g. polygons) which define the photonic device geometry. When combined with volumetric boundary-smoothing techniques \cite{michaels_leveraging_2018, farjadpour_improving_2006, oskooi_accurate_2009}, the quantities $\partial \varepsilon_j/\partial v_i$ and $\partial \mu_j/\partial v_i$ from Eq.\,\ref{eq:1}, which we will refer to collectively as the material Jacobians, are well-defined. 

However, owing to the use of discontinuous boundary representations and geometrical logic operations, $\varepsilon$ and $\mu$ do not have closed-form representations with respect to $\mathbf{v}$. Consequently, photonic simulation and optimization packages such as EMopt (open source) \cite{michaels_leveraging_2018} and LumOpt (maintained by Ansys Lumerical) calculate the material Jacobians using finite-difference, requiring $n+1$ computations of $\varepsilon,\mu$ (which are computed along all voxels within some design region, including half-steps needed for each polarization along the Yee cell).

While this is not a computationally-limiting step for most photonic inverse design problems, calculating the material Jacobians can become a computational burden in problems with large $n$ or when changes in individual design variables $v_i\in\mathbf{v}$ map to global changes along the simulation grid (e.g., $\partial \varepsilon_j/\partial v_i$ is non-zero for a large number of grid elements $j=1,...,m$, exacerbated in large 3D simulation domains). 
For a recent example, please see our prior work in Ref. \cite{sun_adjoint_2023}, where solving Eq.\,\ref{eq:1} using finite-difference ($\sim$30 minutes per iteration on a 144 MPI process calculation, $n=325$, EMopt\,v2019.5.6) constitutes over 50\% of the total optimization wall time, exceeding the combined time of performing the forward and adjoint simulations. We propose to alleviate this problem by using automatic differentiation to efficiently compute Eq.\,\ref{eq:1}.%, requiring only 1 computation of the material distributions $\varepsilon, \mu$ regardless of the number of parameters $n$.

Automatic differentiation (AutoDiff or AD) is an efficient method to calculate exact numerical gradients, often used in machine learning (ML) and deep learning (DL) especially in the context of backpropagation \cite{rumelhart_learning_1986, paszke2019pytorch}. AutoDiff generally can be thought of as a numerical application of the chain rule of Calculus, where the gradient of a composition of analytic functions can be found efficiently by knowing the analytic form of the gradient of each individual function or transform in the composition. The key advantage of modern AutoDiff software packages like PyTorch, TensorFlow, and JAX lies in their ability to automatically track the computational graph associated with this differentiable function composition, enabling scalable gradient computations over large systems of functions or transforms like that in a neural network without the need for a user to manually compute gradients or cache data in memory.

In this paper we propose shape optimization geometry representations that are compatible with automatic differentiation. In other words, we assume that $\varepsilon$ and $\mu$ can be expressed as a composition of differentiable functions or transforms with respect to $\mathbf{v}$:
\begin{align}
    \label{eq:eps}\varepsilon &:= \varepsilon(\mathbf{v}) =  p_P\circ p_{P-1} \circ \cdots \circ p_2 \circ p_1 (\mathbf{v})\,,\\
    \label{eq:mu}\mu &:= \mu(\mathbf{v}) =  q_Q \circ q_{Q-1} \circ \cdots \circ q_2 \circ q_1 (\mathbf{v})\,,
\end{align}
where $p_{j\in\{1,...,P\}}$ and $q_{j\in\{1,...,Q\}}$ are sequences of differentiable functions; without loss of generality, we assume each function is vector-input and vector-valued. In this section, we will maintain this notation for the sake of generality; in the next section we will show how to construct practical geometrical primitives and geometrical compositions of primitives in this way.

Given differentiable representations of the material distributions in Eq.\,\ref{eq:eps}--\ref{eq:mu}, we may immediately express the gradient of the objective %$\frac{\partial f}{\partial v_i}$ in the form of a reverse-mode automatic differentiation:
$\nabla_\mathbf{v} f$ in the form of a reverse-mode automatic differentiation:
%\begin{align}
%    \label{eq:AD}\frac{\partial f}{\partial v_i}=2\omega \textrm{Im}\left(\frac{\partial p_1}{\partial v_i}\bm{\eta}_1-\frac{\partial q_1}{\partial v_i}\bm{\zeta}_1\right),\quad i=1,...,n
%\end{align}
\begin{align}
    \label{eq:AD}\nabla_\mathbf{v} f=2\omega \textrm{Im}\left(\bm{\eta}_0-\bm{\zeta}_0\right)\,,
\end{align}
where $\bm{\eta}_0,\bm{\zeta}_0$ defined as:
%\begin{align}
%    \label{eq:jaceps}\eta^{l-1} = \frac{\partial p_{l}}{\partial p_{l-1}}\eta^l,\quad l=P,...,2, \quad \eta^P=\mathbf{E}\odot\mathbf{E}^\textrm{adj}\\
%    \label{eq:jacmu}\zeta^{l-1} =  \frac{\partial q_{l}}{\partial q_{l-1}}\zeta^l, \quad l=Q,...,2, \quad \zeta^Q=\mathbf{H}\odot\mathbf{H}^\textrm{adj}
%\end{align}
\begin{align}
    \label{eq:jaceps}&\textrm{For }l=P,...,1:\quad\bm{\eta}_{l-1} = \frac{\partial p_{l}}{\partial \bm{a}_l}\bm{\eta}_l, \quad \bm{a}_l=p_{l-1}(\bm{a}_{l-1}), \quad \bm{a}_1=\mathbf{v}, \quad \bm{\eta}_P=\mathbf{E}\odot\mathbf{E}^\textrm{adj}\,,\\
    \label{eq:jacmu}&\textrm{For }l=Q,...,1:\quad\bm{\zeta}_{l-1} =  \frac{\partial q_{l}}{\partial \bm{b}_l}\bm{\zeta}_l, \quad \bm{b}_l = q_{l-1}(\bm{b}_{l-1}),\quad \bm{b}_1=\mathbf{v}, \quad \bm{\zeta}_Q=\mathbf{H}\odot\mathbf{H}^\textrm{adj}\,,
\end{align}
are the Jacobian-vector products, which are computed recursively beginning with the Hadamard (element-wise) product of the vectorized forward and adjoint field terms ($\mathbf{E}\odot\mathbf{E}^\textrm{adj}$ and $\mathbf{H}\odot\mathbf{H}^\textrm{adj}$). There are a few key things to notice here: (i) computing $\varepsilon$ and $\mu$ using Eq.\,\ref{eq:eps}--\ref{eq:mu} consists of $P$ and $Q$ sequential steps, respectively. Similarly, computing the Jacobian-vector products in Eq.\,\ref{eq:jaceps}--\ref{eq:jacmu} consists of $P$ and $Q$ sequential steps, respectively. By comparison, computing the material Jacobians explicitly by finite-difference as in Eq.\,\ref{eq:1} would require $P$ and $Q$ steps, run $n$ times each.\footnote{We have implicitly assumed that $\varepsilon$, $\mu$ are re-evaluated along all $m$ voxels in the simulation domain. While $\varepsilon,\mu$ need to be re-computed $n$ times in the finite-difference scheme, restricting the calculation to non-zero material Jacobian voxels in a principled and efficient way would offer a compelling alternative to AutoDiff.} Furthermore, the finite-difference scheme requires the additional step of storing the material Jacobians in memory and manually sum-reducing them with the forward and adjoint fields as in Eq.\,\ref{eq:1}. (ii) We never explicitly compute the material Jacobians, $\partial \varepsilon_j/\partial v_i$, $\partial \mu_j/\partial v_i$ from Eq.\,\ref{eq:1} when using Eq.\,\ref{eq:AD}--\ref{eq:jacmu}, only the Jacobian-vector products. %(iii) $\bm{\eta}_0$ and $\bm{\zeta}_0$ can also be written $\nabla_\mathbf{v}p_1\bm{\eta}_1$ and $\nabla_\mathbf{v}q_1\bm{\zeta}_1$ -- providing more clarity 
(iii) %The number of multiply-accumulate (MAC) operations needed to compute the Jacobian-vector products in 
Reverse-mode AutoDiff scales most favorably when projection of the design vector to the simulation grid shape (e.g., from $\mathbf{v}\in\mathbb{R}^n$ to $\varepsilon,\mu\in\mathbb{R}^m$, where typically $m\gg n$) occurs late in the composition of functions in Eq.\,\ref{eq:eps}--\ref{eq:mu}, because otherwise more steps of the Jacobian-vector products in Eq.\,\ref{eq:jaceps}--\ref{eq:jacmu} will require more memory and large matrix-vector multiplications.\footnote{If one uses simple analytic functions applied element-wise, the Jacobian-vector product is just a Hadamard product of two vectors, which will scale more favorably than the general case.} 

Interestingly, it is often found in practice that computing $\nabla_\mathbf{v}f$ using Eq.\,\ref{eq:AD} can be nearly as fast as computing $\varepsilon(\mathbf{v})$, $\mu(\mathbf{v})$ via Eq.\,\ref{eq:eps}--\ref{eq:mu}, regardless of the number of parameters $n$, so long as one sticks to using common feedforward operations like matrix multiplication, broadcasting, and per-sample analytic functions. We will show that, compared to computing Eq.\,\ref{eq:1} by finite-difference of the material Jacobians, the scaling of reverse-mode automatic differentiation is negligible with $n$.

Before introducing general geometry representations for shape optimization, we will show how the bi-level density-based topology optimization adjoint variables method gradient can be interpreted as a special application of the reverse-mode AutoDiff gradient (Eq.\,\ref{eq:AD}--\ref{eq:jacmu}) when $\varepsilon,\mu$ are generated in a manner compatible with AutoDiff as in Eq.\,\ref{eq:eps}--\ref{eq:mu}.

\subsection{Bi-level topology optimization gradient computed by automatic differentiation}
\label{sec:topology}
Density-based topology optimization is a popular inverse design geometry representation, where the permittivity/permeability of each simulation grid element in some user-defined region is allowed to vary continuously as an independent degree-of-freedom. However, totally unconstrained topology optimization is rarely practical because we are typically limited to platforms with discrete material properties. Thus, topology optimization is more commonly implemented with bounds on the allowed material values, which is the instructive case we consider here.

For simplicity we will assume that $\mu$ is constant.  Define $\mathbf{v}\in\mathbb{R}^n$ where $n\leq m$ is a vector of parameters that are directly mapped to $m$ permittivity values in some designable region of the simulation grid using the following:
\begin{align}
    \varepsilon(\mathbf{v}) = \mathbf{B}\left(\Delta \varepsilon \sigma(\mathbf{v})+\varepsilon_l\mathbf{1}_n\right)\,,
    \label{eq:TO_example}
\end{align}
where $\sigma:\mathbb{R}\rightarrow[0,1]$ is a smooth, bounded nonlinearity applied per-entry of $\mathbf{v}$ such that $\sigma(\mathbf{v})=[\sigma(v_1),...,\sigma(v_n)]^T$. In this subsection we will assume that $\sigma$ is the sigmoid function. $\mathbf{1}_n$ is an $n$-vector of ones. $\varepsilon_l, \Delta \varepsilon\in\mathbb{R}$ define the lower and upper bounds of permittivity values (e.g., $\varepsilon_l = \varepsilon_{\textrm{SiO}_2}$ and $\Delta\varepsilon = \varepsilon_\textrm{Si} - \varepsilon_{\textrm{SiO}_2}$). $\mathbf{B}\in\mathbb{R}^{m\times n}$ is a fixed broadcasting matrix which might simply be the identity or may enforce, e.g., planarity of integrated photonic devices (since typically we are constrained to 2D lithographic patterning on SOI wafers). In practice, typically $m\geq n\geq m^{2/3}$, and $m$ may often be larger than 100,000. Consequently, computing Eq.\,\ref{eq:1} by finite-difference of $\partial \varepsilon_j/\partial v_i, \forall i=1,...,n; j=1,...,m$ is inefficient and intractable.\footnote{This is not a completely accurate claim in the context of topology optimization, since we may exploit the fact that all $\varepsilon,\mu$ outputs are independent with respect to the parameter inputs $\mathbf{v}$. In particular, we could approximate the full Jacobian in one finite-difference step via $\frac{\partial \varepsilon}{\partial\mathbf{v}}\approx \mathbf{B}\,\textrm{diag}\left\{\Delta \varepsilon\frac{\sigma(\mathbf{v}+h\mathbf{1}_n)-\sigma(\mathbf{v})}{h}\right\}$ for $h>0$. The resulting $m\times n$ matrix is large and inefficient to save in memory if represented densely, but the point stands that it is computed at similar complexity to the corresponding part of the reverse-mode AutoDiff expression shown in Eq.\,\ref{eq:top_AD}. Therefore, the point to emphasize here is that, in general shape optimization scenarios, outputs are not independent with respect to parameters, and therefore finite-difference requires us to compute the Jacobian with respect to each parameter $v_i\in\mathbf{v}$ individually: $\frac{\partial\varepsilon}{\partial \mathbf{v}}\approx[...,
\frac{\partial\varepsilon}{\partial v_i}, ...]$. We use topology optimization as an obvious example where application of finite-differences in this manner is computationally wasteful.}

Instead, we may express Eq.\,\ref{eq:TO_example} as a sequence of 3 function compositions, in the vein of Eq.\,\ref{eq:eps}:
\begin{align}
    \varepsilon(\mathbf{v}) = p_3\circ p_2\circ p_1(\mathbf{v}),\quad
    p_1(\mathbf{a}_1)=\sigma(\mathbf{a}_1),\quad p_2(\mathbf{a}_2)=\Delta \varepsilon \mathbf{a}_2 + \varepsilon_l\mathbf{1}_n, \quad p_3(\mathbf{a}_3)=\mathbf{B}\mathbf{a}_3,
\end{align}
with corresponding Jacobians:
\begin{align}
    \frac{\partial p_3}{\partial \mathbf{a}_3} = \mathbf{B}^T, \quad \frac{\partial p_2}{\partial \mathbf{a}_2} = \Delta\varepsilon\mathbf{I}_{n\times n}, \quad \frac{\partial p_1}{\partial \mathbf{a}_1} = \textrm{diag}\left\{\sigma(\mathbf{a}_1)\odot(\mathbf{1}_n-\sigma(\mathbf{a}_1))\right\},
\end{align}
where we used the identity $\sigma'(\cdot)=\sigma(\cdot)(1-\sigma(\cdot))$ for sigmoid derivatives. The total gradient given in terms of Jacobian-vector products via Eq.\,\ref{eq:AD} may now be expressed as (after some minor simplification):
%\begin{align}
%    \frac{d f}{d \mathbf{v}} =  2\omega\Delta\varepsilon\textrm{diag}\left\{\sigma(\mathbf{v})\odot(\mathbf{1}_n-\sigma(\mathbf{v}))\right\} \mathbf{B}^T \textrm{Im}(\mathbf{E}\odot \mathbf{E}^\textrm{adj})
%\end{align}
\begin{align}
    \label{eq:top_AD}
    \nabla_\mathbf{v} f =  2\omega\Delta\varepsilon\sigma(\mathbf{v})\odot(\mathbf{1}_n-\sigma(\mathbf{v}))\odot \mathbf{B}^T \textrm{Im}(\mathbf{E}\odot \mathbf{E}^\textrm{adj}).
\end{align}
Notably, computing $\nabla_\mathbf{v}f$ is extremely efficient, since we may re-use the calculation of $\sigma(\mathbf{v})$ from the forward pass, and $\mathbf{B}^T\textrm{Im}(\mathbf{E}\odot\mathbf{E}^\textrm{adj})$ typically amounts to a simple sum-reduction over pixel field values. If we choose to remove the constraint on pixel bounds ($p_1(\mathbf{v})=\mathbf{v}$) and let $n=m$ such that $\mathbf{B}=\mathbf{I}_{n\times n}$, then $\nabla_\mathbf{v} f=2\omega\Delta\varepsilon\textrm{Im}(\mathbf{E}\odot\mathbf{E}^\textrm{adj})$ reduces to the well-known, unconstrained topology optimization adjoint method gradient from previous derivations \cite{Miller:EECS-2012-115, lalau-keraly_adjoint_2013}.

\section{AutoDiffGeo: AutoDiff-compatible, smooth geometry primitive representations}
\label{sec:diffgeo}

In this section we describe smooth, elementary shape geometry primitive representations that are compatible with automatic differentiation, which we will abbreviate as AutoDiffGeo. For instance, AutoDiffGeo allows one to invoke a differentiable graph for the permittivity and permeability similar to Eq.\,\ref{eq:eps}-\ref{eq:mu}, such that AutoDiff may be used to compute the adjoint method gradient, Eq.\,\ref{eq:AD}-\ref{eq:jacmu}. All AutoDiffGeo shapes in this paper use the same basic building-block: a (piecewise-)smooth, monotonic, [0,1]-bounded, nonlinear function defined on $\mathbb{R}$. In the following we will refer to this function generally as $\sigma:\mathbb{R}\rightarrow [0,1]$, and, without loss-of-generality, extend this to a dimensionless function $\sigma_k:\mathbb{R}\rightarrow [0,1]$ such that $\sigma_k(x)\mapsto\sigma(k x)$ where $x$ is a coordinate and $k>0$ is a characteristic inverse-length-scale that will be discussed shortly.  Perhaps the most common nonlinearity of this type is the sigmoid function, $\sigma^\text{sigmoid}_k(x):=(1+\exp(-kx))^{-1}$ which asymptotically approaches 0 or 1 in the limit of large negative $x$ or large positive $x$, respectively, with smooth transition about $x=0$. 

\begin{figure}
    \centering
    \includegraphics[width=0.75\textwidth]{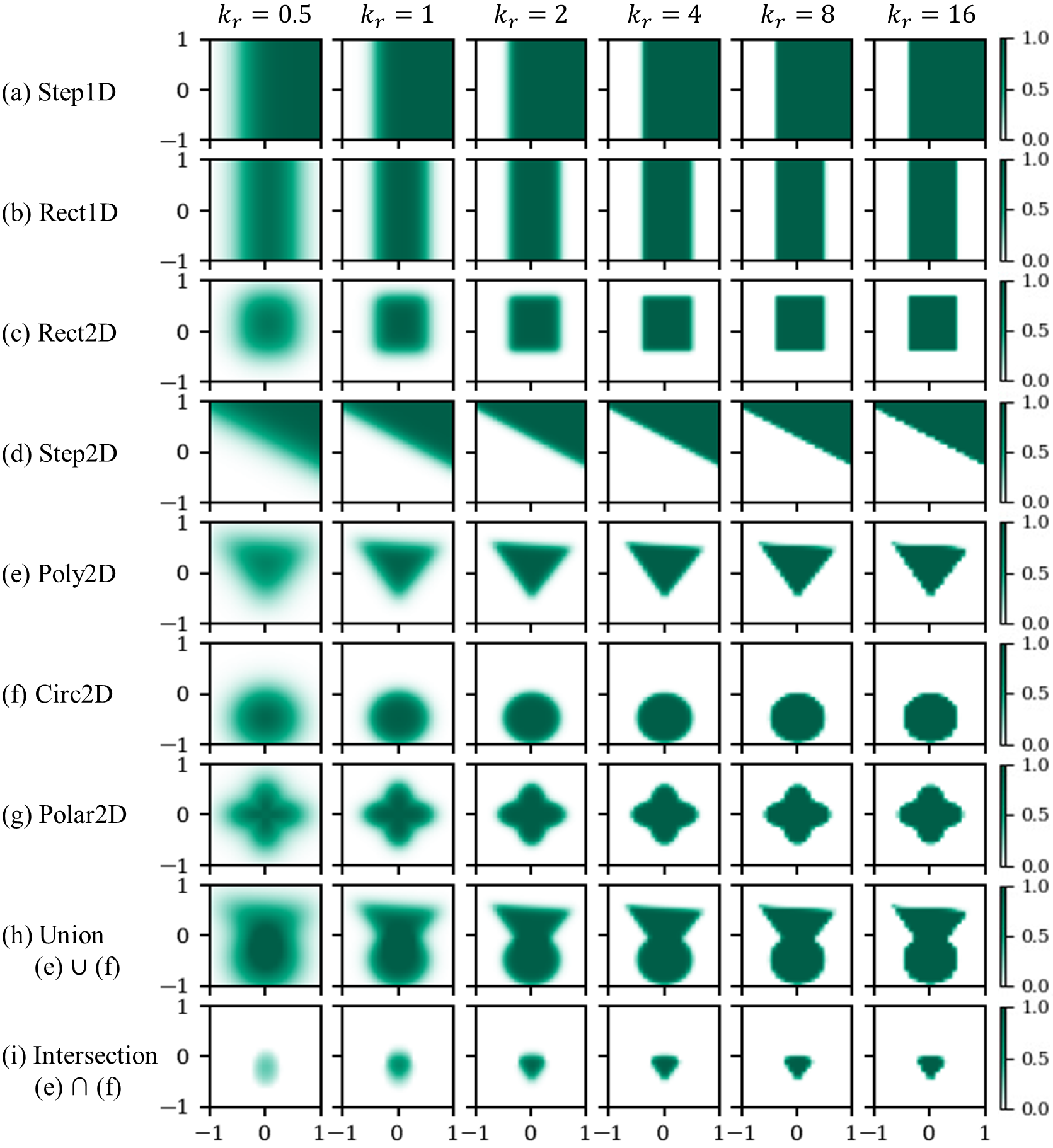}
    \caption{Two-dimensional representations of shape primitives constructed from differentiable function compositions. Here, we assume a sigmoid nonlinearity with characteristic inverse-length-scale $k=k_r/\Delta x$, such that $\sigma_k(x)\mapsto\sigma(k x)$ where $\sigma$ is the sigmoid function and $\Delta x=0.08$ is the uniform mesh discretization. We vary $k_r$ along each column. Function definitions and parameters used in this example may be found in Table\,\ref{tab:autodiffgeo}.}
    \label{fig:shape_examples}
\end{figure}

$\sigma_k$ can be easily extended to a step function with parametrically-defined transition coordinate, $x_0$: $\textrm{Step1D}(x, x_0) = \sigma_k(x-x_0)$.
An example of the Step1D function with $x_0=-0.4$ is depicted in 2D in the first row of Fig.\,\ref{fig:shape_examples}(a). Geometrically-speaking, in 2D, the x-coordinate $x_0$ can be thought of as the boundary between two half-planes with associated material values 0 and 1 on either side of the boundary, and some mix of those material values near the boundary. Throughout Fig.\,\ref{fig:shape_examples}, we use $k=k_r/\Delta x$ where $\Delta x=0.08$ is the underlying uniform grid discretization and $k_r$ is varied along the columns. $k_r$ determines the smoothness of the step transition. For small $k_r$, the transition appears ``blurry'' while for large $k_r$ the transition approaches a ``staircasing'' limit with respect to the finite grid discretization.

In Fig.\,\ref{fig:shape_examples}(b)-(g) we show 6 additional examples of shape primitives that can be built using $\sigma_k$ as an underlying building block. Function definitions and parameters used in these examples may be found in Appendix\,\ref{app:shape_defs}, Table\,\ref{tab:autodiffgeo}. There are 4 items of note: (i) the analysis in Sec.\,\ref{sec:autodiff} suggests that reverse-mode automatic differentiation is made more computationally efficient when projection/broadcasting to multiple spatial coordinates occurs late in the generation of the complete map of permittivity and permeability on the simulation grid. Consequently, the ``2D''-labeled function definitions are less efficient than ``1D''-labeled functions in general with respect to invoking AutoDiff, but are more geometrically expressive. (ii)  The Poly2D function currently only supports convex shapes, and Polar2D supports star-convex shapes (all convex and certain concave shapes). %However, concave polygons can be constructed using the union and intersection operations discussed below.
For other non-star-convex shapes, one will need either the union/intersection functions (discussed below) or a Schwarz-Christoffel conformal mapping \cite{driscoll_schwarz-christoffel_2002}. (iii) Our choices of primitives in Fig.\,\ref{fig:shape_examples}(a)-(g) support additional parameterized transformations not directly depicted here, such as aspect ratio and rotation. (iv) As evident in Fig.\,\ref{fig:shape_examples}(a)-(g) when $k_r$ is small, the boundary transition of shapes may be too gradual, and consequently the desired upper bound of 1 for the material value on the interior of the shape is not attained. This should be taken into consideration for choice of $k_r$ if the user defines shapes that are small relative to the grid spacing $\Delta x$. 

Perhaps our most important contribution in this section is indicated in Fig.\,\ref{fig:shape_examples}(h)-(i), where we show examples of effective Boolean logic in the form of differentiable union and intersection operations. These operations are the key to convenient shape optimization implementations, because it allows one to build arbitrary polygons, convex or concave, from simple primitives as previously illustrated in Fig.\,\ref{fig:ASO_example}. Suppose we are given shapes $\{s_1,...,s_N\}$, each [0,1]-bounded and defined on the same coordinate grid (e.g., $s_i\in[0,1]^m$ where $m=X\times Y\times Z$ is the shape of the local designable region of the grid where shapes $s_1,...s_N$ vary). Then the union and intersection operations may be written:
\begin{align}
    %\textrm{union}(s_1, ..., s_N) &= \min\left\{1,\,\max\left\{0, \sum_{i=1}^N s_i\right\}\right\} \label{eq:union} \\
    \textrm{union}(s_1, ..., s_N) &= \min\left\{1,\,\sum_{i=1}^N s_i\right\} \label{eq:union} \\
    %\textrm{intersection}(s_1, ..., s_N) &= \min\left\{2, \max\left\{1,\,\sum_{i=1}^N s_i\right\}\right\} - 1 \label{eq:intersection}
    \textrm{intersection}(s_1, ..., s_N) &= \max\left\{N-1,\,\sum_{i=1}^N s_i\right\} - (N-1) \label{eq:intersection}
\end{align}
The intuition for Eq.\,\ref{eq:union} is straightforward: given shapes with [0,1]-bounded material values, we simply take their linear superposition element-wise. If for any given grid element the superposition exceeds 1, then we know that one or more shapes occupy that grid element, so we may simply clamp the superposition. Eq.\,\ref{eq:intersection} is implemented similarly, except in this situation we know that the upper bound for the superposition of $N$ shapes is $N$, so the intersection only turns on when $>N-1$ shapes superpose upon any given simulation grid element. We then re-normalize to the [0,1] bounds.

Note that both Eq.\,\ref{eq:union} and Eq.\,\ref{eq:intersection} are continuous, but only piecewise-differentiable functions. Both functions are compatible with AutoDiff, but the transition from partially-overlapping shapes to fully-overlapping or fully-non-overlapping shapes is abrupt. This can cause some instability in gradient descent with large learning rates, since some of the material partial derivatives on the boundaries between overlapping shapes will be 0. Nevertheless, because smoothness is built into the individual shape primitives, and because in the total calculation of the gradient Eq.\,\ref{eq:AD}--\ref{eq:jacmu} we reduce over all grid elements, the union and intersection operations still tend to work well in practice. Furthermore, Eq.\,\ref{eq:union}-\ref{eq:intersection} have the closest resemblance to what one might envision as a truth table defined on continuous scale. We discuss this at length, with comparison to alternatives, in Appendix\,\ref{app:union_truthtable}. Finally, we note that Eq.\,\ref{eq:union} and Eq.\,\ref{eq:intersection} require that the overlapping shapes in the set $\{s_1,...,s_N\}$ and background consist of at most two levels of associated material values. In order to consider overlapping structures of non-binary material systems in the general shape optimization case, one must perform some additional bookkeeping, and apply the union/intersection operations iteratively to sets of shape objects with different material values. We demonstrate union of structures consisting of non-binary material levels in Appendix.\,\ref{app:union_nonbinary}.

\begin{figure}
    \centering
    \includegraphics[width=\textwidth]{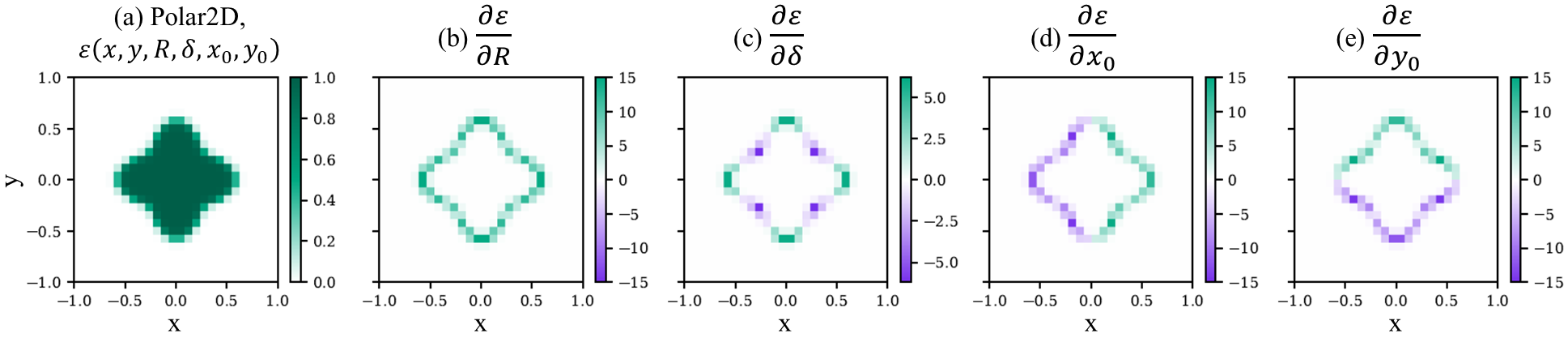}
    \caption{Jacobian of example Polar2D shape assuming sigmoid nonlinearity $\sigma_k$ with characteristic inverse-length-scale $k=4/\Delta x$ where $\Delta x = 0.08$, computed with PyTorch Autograd. The function definition and parameter values $(R, \delta, x_0, y_0)$ are provided in Appendix\,\ref{app:shape_defs}, Table\,\ref{tab:autodiffgeo}.}
    \label{fig:jacobian_example}
\end{figure}

As a matter of completeness, in Fig.\,\ref{fig:jacobian_example} we demonstrate the differentiability of the AutoDiffGeo shape primitives by example, where we plot the Jacobian of the Polar2D example from Fig.\,\ref{fig:shape_examples}(g) with $k=4/\Delta x$, $\Delta x=0.08$, computed with PyTorch Autograd \cite{paszke2019pytorch}. In particular, Fig.\,\ref{fig:jacobian_example}(a) plots the shape $\varepsilon(x,y,\mathbf{v})=\textrm{Polar2D}(x,y,R,\delta,x_0,y_0)$, where $x$ and $y$ are the grid coordinates and $\mathbf{v}=[R,\delta,x_0,y_0]$ are the shape parameters (see Appendix\,\ref{app:shape_defs} for full definition). Fig.\,\ref{fig:jacobian_example}(b)-(e) show the derivatives of each grid element of $\varepsilon$ with respect to each parameter. Derivatives are non-zero approximately only along boundary grid elements of the shape, since only those elements will change meaningfully with respect to an infinitesimal parameter perturbation. We emphasize here that directly computing the material Jacobian $\partial \varepsilon_j/\partial v_i$, is both unnecessary and inefficient with regard to calculating the adjoint method gradient using reverse-mode automatic differentiation in Eq.\,\ref{eq:AD}--Eq.\,\ref{eq:jacmu}, since these equations only require the Jacobian-vector products. Nevertheless, it is included here for purposes of instruction.

\section{Improving discretization error of AutoDiffGeo shape representations and optimizations on rectilinear simulation grids}
\label{sec:subpixelsmoothing}
Up to this point we have introduced (1) how to accelerate adjoint method based inverse design optimization using automatic differentiation, and (2) smooth, closed-form shape primitive representations that are compatible with automatic differentiation.  We have been agnostic as to whether or when these shape representations are physically meaningful as represented on simulation grids. Indeed, we introduce error in FDFD and FDTD by discretizing simulation grids and their corresponding material distributions sampled from structural shape objects. Discretization error is fundamental to numerical solutions of partial differential equations, but it can be exacerbated when simulation grid element material values sampled along structural boundaries are not appropriately smoothened. Fortunately, there are several existing methods to improve discretization error of shape representations in FDFD/FDTD, by means of sub-pixel smoothing or boundary smoothing algorithms \cite{farjadpour_improving_2006, oskooi_accurate_2009, michaels_leveraging_2018}. How do we ensure that our use of AutoDiffGeo shape representations for improved optimization speed does not further exacerbate discretization error in FDFD/FDTD? In this section, we provide two proposals towards enforcing convergence of AutoDiffGeo and standard sub-pixel smoothing representations of shapes in optimized inverse design solutions, summarized in Table.\,\ref{tab:prop1and2} and detailed below.

\begin{table}[ht]
\centering
\begin{threeparttable}
	\caption{Summary of proposals for using AutoDiffGeo representations in inverse design optimizations.}
	\begin{tabular}{llll}
		\toprule
             Version & Simulation Geometry & Gradient Calculation & Effect on Optimization \\
		\midrule
            Standard (FD) & Standard $\varepsilon,\mu$ & Standard $\varepsilon,\mu$ & Exact $\mathbf{E},\mathbf{H}$; Exact $\nabla_\mathbf{v}f$; Slow $\nabla_\mathbf{v}f$ \\
            Prop. 1 (AD) & AutoDiffGeo $\varepsilon,\mu$ & AutoDiffGeo $\varepsilon,\mu$ & *Approximate $\mathbf{E},\mathbf{H}$; Exact $\nabla_\mathbf{v}f$; Fast $\nabla_\mathbf{v}f$ \\
            Prop. 2 (AD) & Standard $\varepsilon,\mu$ & AutoDiffGeo $\varepsilon,\mu$ & Exact $\mathbf{E},\mathbf{H}$; Approximate $\nabla_\mathbf{v}f$; Fast $\nabla_\mathbf{v}f$  \\
		\bottomrule
	\end{tabular}
    \centering
    \begin{tablenotes}
        \item * $\nabla_\mathbf{v} f$ is exact with respect to the approximate $\mathbf{E},\mathbf{H}$ used to compute the objective value, $f(\mathbf{E},\mathbf{H})$. In some cases, the $\mathbf{E},\mathbf{H}$ fields resulting from AutoDiffGeo geometries can be numerically equal to the Standard case. 
        \item For calculation of shape sensitivities: FD = Finite-Difference, AD = Automatic Differentiation.
    \end{tablenotes}
    \label{tab:prop1and2}
\end{threeparttable}
\end{table}

\paragraph{Prop. 1: optimize with AutoDiffGeo shapes now, then enforce strict sub-pixel smoothing}
In this proposal, we directly define the material distributions within the discretized simulation design region using the AutoDiffGeo shapes introduced in Sec.\,\ref{sec:diffgeo}. We use these representations in the solution to the forward and adjoint simulations, and for the gradient calculation in Eq.\,\ref{eq:AD}-\ref{eq:jacmu}. The advantage of this approach is that the gradient calculation is exact, since $\mathbf{E}, \mathbf{H}, \mathbf{E}^\textrm{adj}, \mathbf{H}^\textrm{adj}$ are all computed with respect to the same FDFD/FDTD grid as the defined material distributions $\varepsilon, \mu$. The disadvantage of this approach is that the shape representations themselves in general do not obey exact sub-pixel smoothing, e.g., along dielectric boundaries, and therefore introduce additional discretization error into the field solution with respect to the true electromagnetic solution. This requires us to enforce physical viability of our structural representations after an initial AutoDiff-enhanced optimization is complete, by converting the optimized AutoDiffGeo structure to an exact representation that obeys sub-pixel smoothing. To minimize the penalty of this conversion, one should carefully choose parameters of the AutoDiffGeo shape representation to minimize the difference between it and the exact representation. We will show that some AutoDiffGeo shapes and transformations can be defined with zero error in this approximation.

\paragraph{Prop. 2: enforce strict sub-pixel smoothing now, but use AutoDiff-enhanced computation of the gradient for optimization (with noise)}
In this proposal, we suggest defining the simulation grid with standard, exact shape representations for the purposes of computing $\mathbf{E}, \mathbf{H}, \mathbf{E}^\textrm{adj}, \mathbf{H}^\textrm{adj}$. We only use the AutoDiffGeo versions of the shape representations in the accelerated calculation of the gradient via Eq.\,\ref{eq:AD}--\ref{eq:jacmu}. The advantage of this approach is that physical viability of shape representations is no longer a concern; the approximation error of any given simulation is limited only by the finite grid discretization. The disadvantage of this approach is that the gradient calculation is generally an inexact, noisy approximation of the true gradient, which could make it difficult to locate sharp local optima. %In practice this is not a strong concern, as stochastic gradient descent optimizers are known to perform well in practice. Nevertheless, additional refinement optimizations may be needed to locate sharp local optima. 
An additional disadvantage is that this approach requires a redundant calculation of the material distributions, once for the standard, exact shape representations and once for the AutoDiffGeo versions. However, the exact version may be leveraged to refine the parameters of the AutoDiffGeo version and minimize overall error in the gradient calculation, as opposed to choosing parameters a priori as in Prop. 1.

The disadvantages of Prop. 1 and Prop. 2 are both directly related to the problem of sub-pixel smoothing on rectilinear simulation grids. We address this below.

\subsection{Approximating sub-pixel smoothing with AutoDiffGeo representations}
\label{sec:subpixelapprox}

Several papers have demonstrated that discretization error is minimized in finite-difference based electromagnetic simulators when material properties along boundaries are volumetrically averaged over grid cells \cite{oskooi_accurate_2009, farjadpour_improving_2006, michaels_leveraging_2018}. This is called boundary smoothing or sub-pixel smoothing. In the most rigorous version of sub-pixel smoothing from perturbation theory, when a shape overlaps a discrete grid cell, the resulting displacement field can be viewed as an anisotropic permittivity tensor acting on the electric field, formulated from both the arithmetic mean and harmonic mean of the permittivity values occupying the grid cell and the orientation of the shape's boundary with respect to the grid cell \cite{farjadpour_improving_2006, oskooi_accurate_2009} -- a numerical analog of the elementary electromagnetic interface conditions. However, here we will compare our AutoDiffGeo representations to only a simple scalar arithmetic average of the materials per grid cell (e.g., $\langle \varepsilon \rangle$) in the flavor of Ref.\,\cite{michaels_leveraging_2018}. This version of sub-pixel smoothing has larger error for a given fixed grid discretization \cite{oskooi_accurate_2009}, but is easier to understand and implement. Nevertheless, we note that the discussion and analysis here can likely be extended to the more general sub-pixel smoothing case via two additional transformations: (1) an approximation of the harmonic average of the materials along each grid cell (e.g., $\langle \varepsilon^{-1}\rangle^{-1}$) as defined by AutoDiffGeo shapes, and (2) a mapping of the output of the arithmetic/harmonic mean approximations to an anisotropic material tensor, leveraging knowledge of the orientation of the shape's boundary with respect to the grid cell. We consider these extensions to be future work.

\begin{figure}
    \centering
    \includegraphics[width=0.7\textwidth]{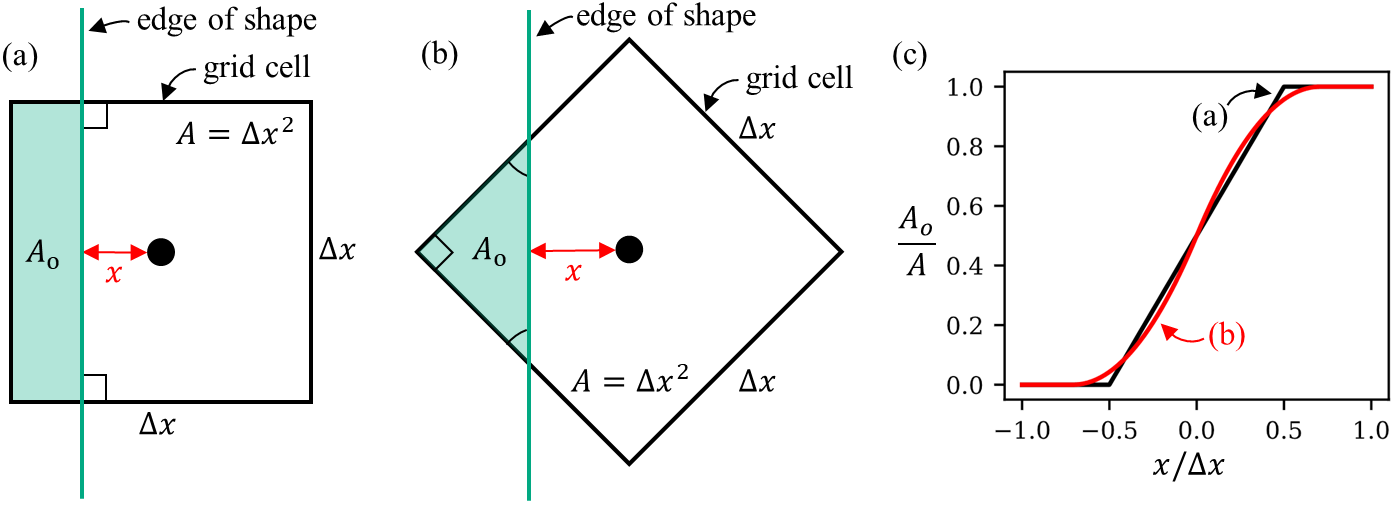}
  \caption{Special case 2D geometrical configurations of a shape with straight edge overlapping a square grid cell with orientation of $0^\circ$ (a) and $45^\circ$ (b). The corresponding normalized area of overlap as a function of $x$ is given in (c).}
  \label{fig:area_nonlinearity}
\end{figure}

In simple geometric configurations, volumetric averaging of material values over grid cells can be posed exactly and analytically. Consider the configurations depicted in Fig.\,\ref{fig:area_nonlinearity}. In both configurations, we consider a shape with straight edge overlapping a square grid cell with side length $\Delta x$. In Fig.\,\ref{fig:area_nonlinearity}(a) the shape's edge is parallel the grid cell's left edge, while in Fig.\,\ref{fig:area_nonlinearity}(b) it is oriented at $45^\circ$. Here, $x$ is the signed-distance of the grid cell center from the edge of the shape. Suppose the material value of the shape and the background is $\varepsilon_s$ and $\varepsilon_b$, respectively. Then the volumetrically-averaged material value of the 2D grid cell can be written:$\langle\varepsilon\rangle=\varepsilon_b + (\varepsilon_s - \varepsilon_b)a_o$
where $a_o:=A_0/A$ is the normalized area of overlap between the shape and the grid cell with $A=\Delta x^2$. We may express $a_o$ as a function of $x$ for the configurations in Fig.\,\ref{fig:area_nonlinearity}(a) and Fig.\,\ref{fig:area_nonlinearity}(b) using the following equations:
\begin{align}
    a_o^\textrm{Fig.\,\ref{fig:area_nonlinearity}(a)}(x) &= \frac{x}{\Delta x}+\frac{1}{2}, \quad -\frac{1}{2}\leq \frac{x}{\Delta x}\leq \frac{1}{2} \label{eq:linear}\\
    a_o^\textrm{Fig.\,\ref{fig:area_nonlinearity}(b)}(x) &= \begin{cases}
        \left(\frac{1}{\sqrt{2}}+\frac{x}{\Delta x}\right)^2, \quad -\frac{1}{\sqrt{2}}\leq \frac{x}{\Delta x} < 0 \\
        1-\left(\frac{1}{\sqrt{2}}-\frac{x}{\Delta x}\right)^2, \quad 0\leq \frac{x}{\Delta x}\leq\frac{1}{\sqrt{2}} \\
    \end{cases} \label{eq:quadratic}
\end{align}

plotted in Fig.\,\ref{fig:area_nonlinearity}(c) for reference.  Unsurprisingly, the forms of Eq.\,\ref{eq:linear} and Eq.\,\ref{eq:quadratic} with respect to $x$ are different, but provide a few interesting insights: (i) Perfect volumetric-averaging of the material values cannot be achieved as a simple function of $x$ without careful consideration of the orientation between shape and grid cell; and while not considered here, aspect ratio ($\Delta x/\Delta y$), non-trivial 3D orientations, and the curvature of the shape's edge within the grid cell will also affect the form of $a_o(x)$. (ii) Both Eq.\,\ref{eq:linear} and Eq.\,\ref{eq:quadratic} are [0,1]-bounded nonlinear functions, similar to the sigmoid function considered earlier, providing a clear physically-motivated connection to the AutoDiffGeo shape representations in Sec.\,\ref{sec:diffgeo} with $\sigma_k(x):=a_o(x)$ where $k=1/\Delta x$.

As a consequence of (i), it is unknown at the time of this paper whether it is possible to construct a parametric $\sigma_k(x)$ nonlinearity that can provide perfect volumetric averaging over all simulation grid cells for arbitrary shape primitives provided in this paper, while maintaining automatic differentiability. Therefore, we take an experimental approach in the choice of simple $\sigma_k(x)$ options for the implementations of Prop. 1 and Prop. 2, detailed below.

\subsection{Experimental choice of nonlinear function for Prop. 1}
\label{sec:exp_prop1}
For Prop.\,1, summarized in Table\,\ref{tab:prop1and2}, we define the FDFD/FDTD simulation grid material distributions, $\varepsilon,\mu$, using AutoDiffGeo shapes. In this subsection, we detail the choice of smooth, nonlinear function $\sigma_k$ needed to define AutoDiffGeo shapes in an effort to minimize error in approximating exact representations of shapes that satisfy volumetric-averaging of material values over grid cells.

As discussed previously, our AutoDiffGeo shapes require the choice of a [0,1]-bounded nonlinearity $\sigma_k$, which determines the smoothness of the shape primitive on the underlying simulation grid. This gives us 2 degrees-of-freedom to improve the sub-pixel smoothing representation of the AutoDiffGeo primitives: the choice of non-linear function $\sigma$ and inverse-length-scale $k$. Several choices of $\sigma_k(\cdot)$ function definitions are provided in Appendix\,\ref{app:nonlinearities} with definitions in Table\,\ref{tab:nonlinearfunctions} and examples plotted in Fig.\,\ref{fig:nonlinearities}. In particular, we include well-known analytic functions such as sigmoid and erf (normalized to [0,1]-bounds) as well as special piecewise differentiable functions abbreviated as sin, linear, and quadratic. $\textrm{linear}(k,x)$ and $\textrm{quadratic}(k,x)$ are generalized versions of the volumetrically-averaged nonlinearities suggested by Eq.\,\ref{eq:linear} and Eq.\,\ref{eq:quadratic} above, but where we allow $k=k_r/\Delta x$ to vary.

Of particular note, all $\sigma_k$ options in Fig.\,\ref{fig:nonlinearities} approach a similar discontinuous step function as $k\rightarrow\infty$ on discrete simulation grids. By choosing a large $k$ satisfying this property, we can obtain exact volumetric averaging (regardless of the choice of base function $\sigma$) by numerical quadrature (super-sampling) the grid cell. Unfortunately, this is more computationally intensive than sampling a single coordinate within the grid cell, and more importantly, incompatible with automatic differentiation (since derivatives of boundary grid elements $\rightarrow \infty$ as $k\rightarrow\infty$). For computational performance, we desire a one-shot-sampling of each grid cell, with purposefully chosen finite-$k$ value. 

\begin{figure}
    \centering
    \includegraphics[width=0.8\textwidth]{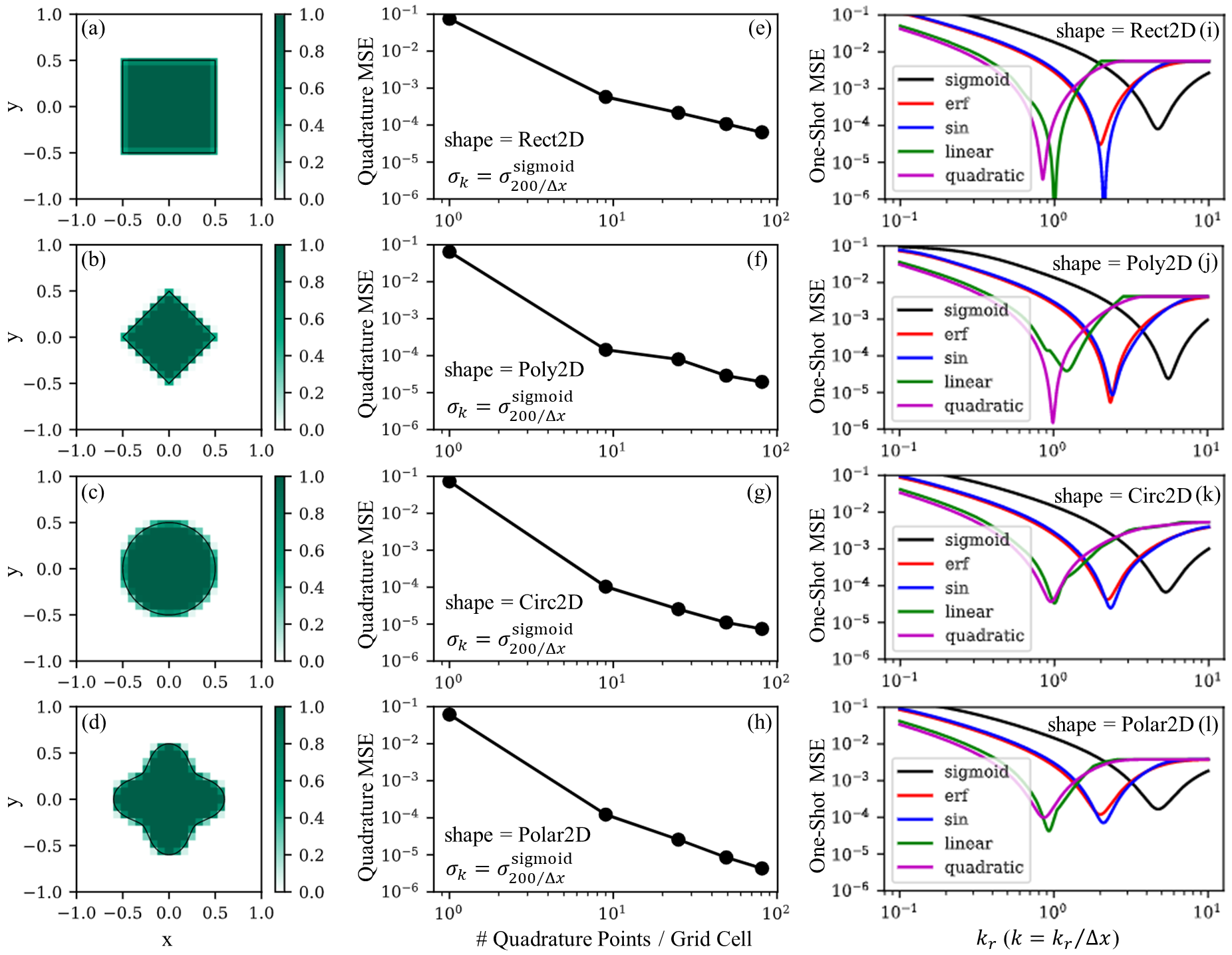}
    \caption{Mean-squared-error (MSE) comparison in approximation of exact volumetrically-averaged shape representations (a)-(d), using numerical quadrature with large-$k$ (e)-(h), and using one-shot sampling with finite-$k$ (i)-(l).}
    \label{fig:quadrature_oneshot}
\end{figure}

Fig.\,\ref{fig:quadrature_oneshot} provides a comparison of numerical quadrature with large $k$ versus one-shot-sampling with finite-$k$ for various choices of nonlinear function, $\sigma_k$. Ground-truth shapes generated via EMopt's exact volumetric-averaging sub-pixel smoothing algorithm, represented as 1000-vertex polygons with $\Delta x = 0.08$ 
uniform mesh discretization, are provided in Fig.\,\ref{fig:quadrature_oneshot}(a)-(d) for reference. We approximate the exact sub-pixel smoothing of these ground-truth shapes using the AutoDiffGeo shape definitions Rect2d, Poly2D, Circ2D, and Polar2D defined in Table\,\ref{tab:autodiffgeo}, originally discussed in Sec.,\ref{sec:diffgeo}. In particular, in Fig.\,\ref{fig:quadrature_oneshot}(e)-(h) we show the mean-squared-error (MSE) of exhaustive numerical quadrature versus the number of quadrature points per grid cell (sampled on a uniform subgrid), where we use a sigmoid function for $\sigma_k$ with large value of $k=200/\Delta x$ such that it approximates a discontinuous step function. In Fig.\,\ref{fig:quadrature_oneshot}(i)-(l) we compare the MSE of several nonlinear functions $\sigma_k$ with varied $k=k_r/\Delta x$ in approximation of the ground-truth shapes using a one-shot-sampling of each grid cell (at cell center). The choice of $\sigma_k$ can have a dramatic effect on the approximation. For the trivial rectangular shapes in Fig.\,\ref{fig:quadrature_oneshot}(a)-(b) we can greatly outperform exhaustive numerical quadrature with the appropriate choice of piecewise sin, linear, or quadratic nonlinearity (in fact, the piecewise linear function approaches zero error within numerical precision for the case in Fig.\,\ref{fig:quadrature_oneshot}(a)). For the nontrivial shapes in Fig.\,\ref{fig:quadrature_oneshot}(c)-(d) we find that piecewise sin, linear, and quadratic functions provide the best overall MSE $\approx2$-$4\times10^{-5}$, indicating that one-shot-sampling can be as effective as exhaustive numerical quadrature with $\sim$9-16 samples per grid cell.

\begin{figure}
    \centering
    \includegraphics[width=0.8\textwidth]{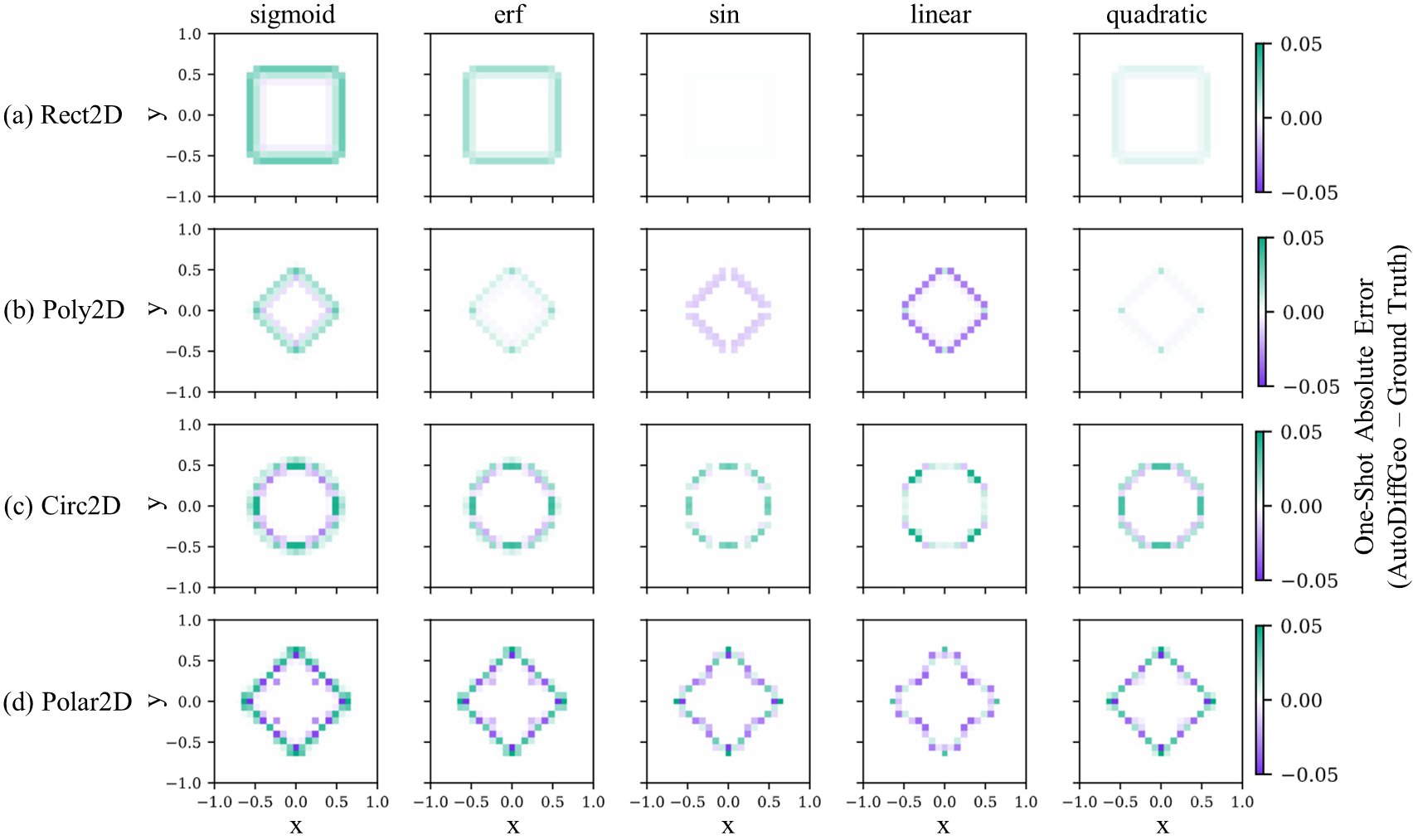}
    \caption{Comparison of boundary errors using one-shot approximation of exact volumetrically-averaged shape representations at optimal $k$.}
    \label{fig:boundary_error}
\end{figure}

The boundary errors of one-shot sampling with each nonlinear function $\sigma_k$ at the optimal $k$ value from Fig.\,\ref{fig:quadrature_oneshot}(i)-(l) are visualized in Fig.\,\ref{fig:boundary_error}. Of particular interest are the boundary errors of the piecewise linear and quadratic functions, which confirm the theoretical discussion in Sec.\,\ref{sec:subpixelapprox} above. In particular, we observe that the linear function yields zero error along grid cells where a polygonal edge is parallel to the grid edge. The quadratic function yields zero error along grid cells where the polygonal edge clips the grid exactly at 45$^\circ$. For the Poly2D shape in Fig.\,\ref{fig:boundary_error}(b), which is a square rotated by 45$^\circ$, the quadratic function yields non-zero error only along the corners of the shape where there are multiple edges overlapping the grid cells.

Meanwhile, the piecewise sin function provides a good balance at all orientations of shape versus grid cell boundaries. These results emphasize that the choice of nonlinear function $\sigma_k$ can vary with application, or even on a per-shape basis. For instance, in shape optimization of structures like the MMI-like device from Fig.\,\ref{fig:ASO_example}, a piecewise linear $\sigma_k$ is preferable; in fact it will provide zero error within numerical precision. In integrated photonics applications where we define non-trivial shapes in the $xy$ plane, but still maintain planarity along the $z$ axis, we may desire to use a piecewise sin or piecewise quadratic function to define features in the $x,y$ coordinate directions, but use a piecewise linear function to define the shapes' depth in the $z$ coordinate direction.

\subsection{Experimental choice of nonlinear function for Prop. 2}
\label{sec:prop2}
For Prop.\,2, summarized in Table\,\ref{tab:prop1and2}, we define the FDFD/FDTD simulation grid with standard, exact shape representations that satisfy volumetrically-averaging over each grid cell for the purpose of computing  $\mathbf{E}, \mathbf{H}, \mathbf{E}^\textrm{adj}, \mathbf{H}^\textrm{adj}$, but then use the AutoDiffGeo versions of the shape representation to compute the reverse-mode AutoDiff gradient of the electromagnetic objective via Eq.\ref{eq:AD}-\ref{eq:jacmu}. Similar to Prop.\,1, we need to choose the appropriate nonlinear function $\sigma_k$ in our instantiations of AutoDiffGeo shapes. However, in this case, our goal is to minimize the error of the AutoDiff gradient (computed using AutoDiffGeo shapes) with respect to the conventional finite-difference (FD) gradient (computed using standard shape definitions with exact volumetric-averaging). This is equivalent to approximating the FD material Jacobian (e.g., $\partial \varepsilon_j/\partial v_i$) with the Jacobian of AutoDiffGeo shapes.

We first establish the analytical, exact volume-weighted material derivative for the simple configurations of shape 
edge and grid cell considered previously in Fig.\,\ref{fig:area_nonlinearity}(a)-(b), where the shape's edge clips the grid cell at $0^\circ$ and $45^\circ$, respectively. Suppose the material value of the interior of the shape and the background is $\varepsilon_s$ and $\varepsilon_b$, respectively. Furthermore, recall that the signed-distance from shape edge to grid cell center is denoted $x$. Using the chain rule, we can decompose the derivative of the material value of grid cell $j$ with respect to shape parameter $i$ as $\frac{\partial\varepsilon_j}{\partial v_i}=\frac{\partial\varepsilon_j}{\partial x_j}\frac{\partial x_j}{\partial v_i}$. Both standard and AutoDiffGeo methods implicitly define equivalent parameterizations of $\frac{\partial x_j}{\partial v_i}$, which represents the perturbation to a shape's boundary with respect to shape parameter $v_i$, measured with respect to each grid cell center $x_j$. Therefore, it suffices to demonstrate expressions for $\frac{\partial \varepsilon_j}{\partial x_j}$. As discussed previously, for the simple 2D configurations in Fig.\,\ref{fig:area_nonlinearity}(a)-(b), the volumetrically-averaged material value can be expressed as $\langle \varepsilon \rangle = \varepsilon_b + (\varepsilon_s - \varepsilon_b)a_o$ where $a_o$ is the normalized area of overlap of the shape with the grid cell. Expressions for $a_o(x)$ were provided in Eq.\,\ref{eq:linear}-\ref{eq:quadratic}. Therefore, $\frac{\partial \langle \varepsilon \rangle}{\partial x}=(\varepsilon_s - \varepsilon_b)\frac{\partial a_o(x)}{\partial x}$, where,
\begin{align}
    \frac{\partial}{\partial x}a_o^\textrm{Fig.\,\ref{fig:area_nonlinearity}(a)}(x) &= \frac{1}{\Delta x}, \quad -\frac{1}{2}\leq \frac{x}{\Delta x}\leq \frac{1}{2} \label{eq:dlinear}\\
    \frac{\partial}{\partial x}a_o^\textrm{Fig.\,\ref{fig:area_nonlinearity}(b)}(x) &= \begin{cases}
        \frac{\sqrt{2}\Delta x + 2x}{\Delta x^2}, \quad -\frac{1}{\sqrt{2}}\leq \frac{x}{\Delta x} < 0 \\
        \frac{\sqrt{2}\Delta x - 2x}{\Delta x^2}, \quad 0\leq \frac{x}{\Delta x}\leq\frac{1}{\sqrt{2}} \\
    \end{cases} \label{eq:dquadratic}
\end{align}
In particular, these analytical expressions imply angular dependence of the derivative of volumetrically-averaged materials even in the overly simplistic cases of Fig.\,\ref{fig:area_nonlinearity}(a)-(b), and it is unknown at the time of this paper whether there exists AutoDiff-compatible nonlinear functions in 2D/3D that can be constructed a priori to yield the exact derivatives for arbitrary shapes and in all directions. However, we have also shown that the derivative of the volumetrically-averaged material value per grid cell can be expressed simply in terms of the derivative of a nonlinear function $\sigma_k$ that approximates $a_o$. Below we experimentally compare various choices of $\sigma_k$ in an effort to improve this approximation. Note that AutoDiff will automatically compute $\partial \sigma_k/\partial x$, so we need not manually define these derivatives ourselves. 

\begin{figure}
    \centering
    \includegraphics[width=0.99\textwidth]{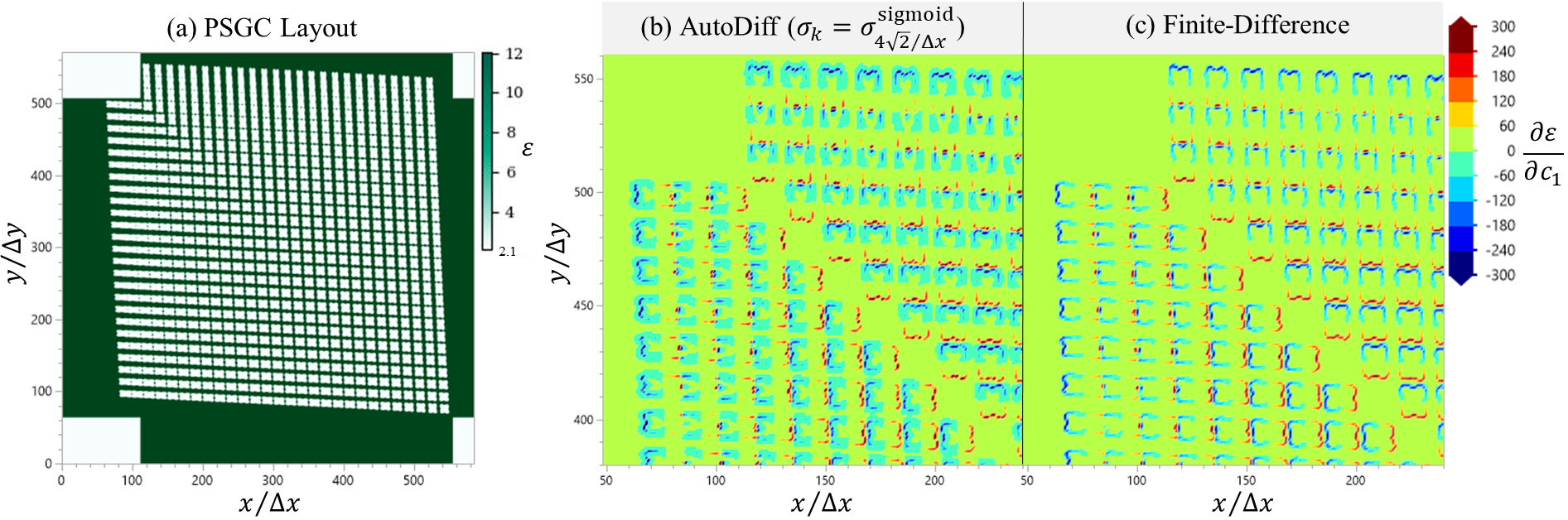}
    \caption{Layout of PSGC expressed in terms of relative permittivity (a), and corresponding partial derivatives of the permittivity with respect to the first order cosine coefficient $c_1$ when using the AutoDiff-method (sigmoid parameter $k=4\sqrt{2}/\Delta x=188.6$) (b) and exact volume-weighted FD method (c). (b) and (c) are zoomed into upper left corner of (a). $\Delta x=\Delta y=0.03\textrm{\textmu m}$.}
    \label{fig:psgc_full_jac_smh}
\end{figure}

We use the example of a polarization-splitting grating coupler (PSGC) \cite{sun_adjoint_2023} layout to study the impact of $\sigma_k$ choices on the errors in the permittivity Jacobian. We will explain the PSGC functionality and optimize its geometry in Sec.\,\ref{sec:PSGC}.  Fig.\,\ref{fig:psgc_full_jac_smh}(a)  shows the top-view layout of the PSGC: the grating area consists of a 2D array of $\sim800$ scatterers, which are etched into a $\textrm{Si}$ slab and backfilled with $\textrm{SiO}_2$. We adopt the parameterization from Ref.\,\cite{sun_adjoint_2023}, where each scatterer is synthesized by sine and cosine basis functions up to order $N=6$ in polar coordinates as in the following equation, where $c_n$ and $s_n$ are the trigonometric coefficients and $R_0$ the DC component:
\begin{align}
    R(\theta)=R_0+\sum_{n=1}^{N}(c_n\cos(n\theta+\pi)+s_n\sin(n\theta+\pi)).
    \label{eq:PSGCboundary}
\end{align}
To generate a standard volumetrically-averaged permittivity layout, we define polygons with boundaries parametrically-defined by Eq.\,\ref{eq:PSGCboundary} using EMopt's \cite{anstmichaels_emopt_2023, michaels_leveraging_2018} grid smoothing module. Meanwhile, we generate the permittivity layout of AutoDiff-compatible scatterers using,
\begin{align}
    \varepsilon_\textrm{scatterer}(r,\theta)=\varepsilon_{\textrm{SiO}_2}+(\varepsilon_\textrm{Si}-\varepsilon_{\textrm{SiO}_2})\sigma_k(R(\theta)-r)\;,
\end{align}
where we choose various nonlinear functions $\sigma_k$ from Table\,\ref{tab:nonlinearfunctions}, including sigmoid, linear, quadratic, erf, and sine function.
As a representative example, the derivatives of the permittivity with respect to the first order cosine coefficient $c_1$ (e.g., $\partial \varepsilon/\partial c_1$ for all simulation voxels) is calculated by the exact volume-weighted FD method and the AutoDiff method, respectively.  Fig.\,\ref{fig:psgc_full_jac_smh}(b)-(c) shows the calculated permittivity derivative maps of the FD method and the AutoDiff method with sigmoid smooth function and $k=4\sqrt{2}/\Delta x=188.6$ on the cross section of the grating area at FDTD grid size of $\Delta x=0.03\textrm{\textmu m}$, which consists of $585 \times 572 =334,620$ grid points. Visually, there is a close resemblance between the two derivative calculation methods, although the AutoDiff method tends to output non-zero derivatives at multiple pixel distance from shape boundaries due to the smoothness of the sigmoid function.

\begin{figure}
    \centering
    \includegraphics[width=0.9\textwidth]{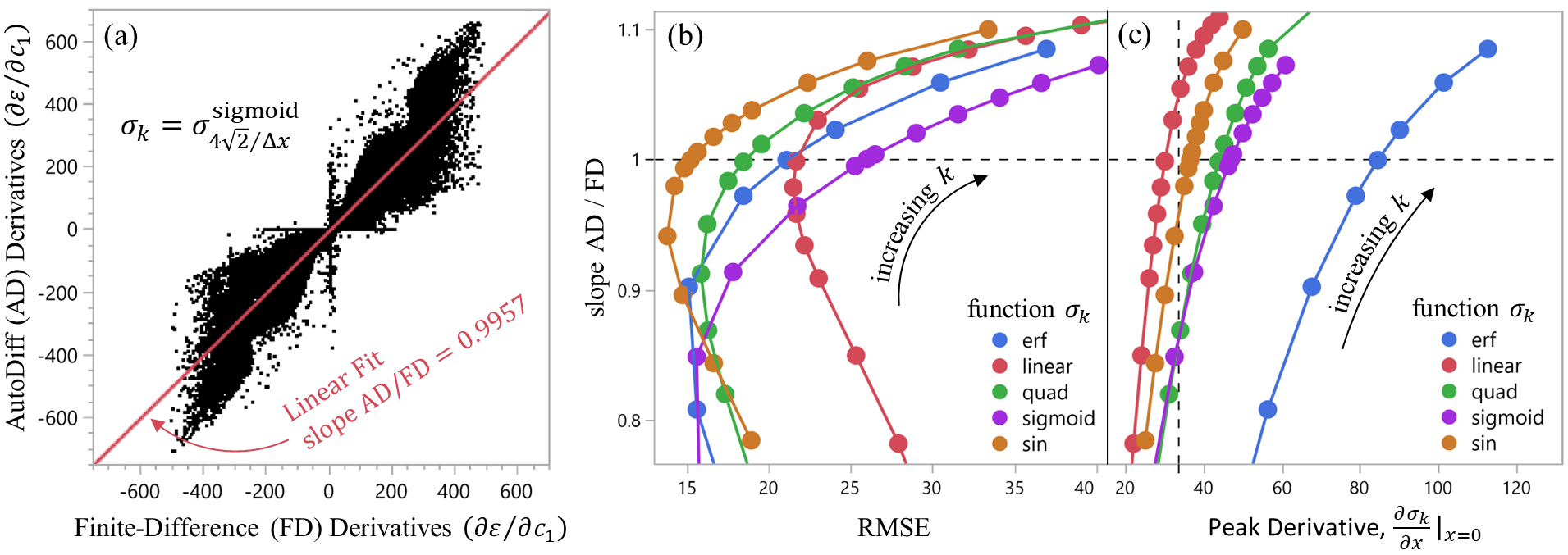}
    \caption{(a). Linear fitting of the AD- vs. FD-gradient of the PSGC, with sigmoid parameter \textit{k}=188.6.  (b) Fitting slope vs. fitting RMSE and \textit{k}/4 of the sigmoid smooth function. }
    \label{fig:psgc_jac_fitting_smh}
\end{figure}

The AutoDiff method and FD-method calculated derivatives are fitted to a linear model with intercept forced to 0, as shown in Fig.\,\ref{fig:psgc_jac_fitting_smh}(a).  Note that simple linear regression is used here in lieu of orthogonal regression, since the FD-method is deemed exact.  The linear fitting slope and root-mean-squared-error (RMSE) are extracted for all smooth functions and various $k$ values, as shown in Fig.\,\ref{fig:psgc_jac_fitting_smh}(b).  The fitting slope increases with the inverse-length-scale parameter $k$, since the peak derivative of all nonlinear functions ($\partial \sigma_k/\partial x$ evaluated at $x=0$) increases monotonically with $k$, as shown in Fig.\,\ref{fig:psgc_jac_fitting_smh}(c). We find that the fitting RMSE is not always minimized at fitting slope=1.0 nor at the same $k$ value due to the different functional forms. Importantly, since RMSE of the permittivity derivatives is never identically zero for any choice of $\sigma_k$, the use of AutoDiff-generated layout for the gradient calculation via Prop.\,2 in optimization will result generally in a noisy approximation of the true gradient. Consequently, in our experiment utilizing Prop.\,2 below (Sec.\,\ref{sec:PSGC}), we will study the effect of using each nonlinear function $\sigma_k$, where $k$ is chosen as the value that minimizes RMSE in Fig.\,\ref{fig:psgc_jac_fitting_smh}(b).

\section{Experiments}
\label{sec:experiments}
In this section we validate the theoretical claims from Sec.\,\ref{sec:autodiff}-\ref{sec:diffgeo} and test our proposals from Table\,\ref{tab:prop1and2} using several shape optimization inverse design applications enhanced with automatic differentiation (AutoDiff or AD) in the gradient calculation, in comparison to standard finite-difference (FD). We show two examples of Prop.\,1 in the blazed grating coupler and non-adiabatic taper optimizations in Sec.\,\ref{sec:blazed_grating}-\ref{sec:taper2D}, and one example of Prop.\,2 in the polarization-splitting grating coupler optimization of Sec.\,\ref{sec:PSGC}. All comparisons use EMopt (version 2023.1.1) \cite{anstmichaels_emopt_2023, michaels_leveraging_2018} as the underlying electromagnetic field simulator, and its corresponding default module for generation of volumetrically-averaged permittivity layout needed for finite-difference calculations of the material Jacobian in the adjoint method gradient.

\subsection{Single-polarization blazed grating coupler in 2D}
\label{sec:blazed_grating}
Grating couplers consist of $N_g$ periodically-arranged scattering elements in a waveguide. Typical single-polarization grating couplers consist of one photolithography / etch step to define scattering elements, while blazed gratings consist of at least two etches with varied depths (e.g., see \cite{watanabe_perpendicular_2017}). To demonstrate the versatility and speed of automatic differentiation in photonic shape optimization, we will compare the AutoDiff-enhanced gradient calculation to standard Finite-Difference in the optimization of a two-etch blazed grating, with explicit scaling of gradient calculation time and optimization wall-clock time versus number of variables.

\begin{figure}
    \centering
    \includegraphics[width=0.9\textwidth]{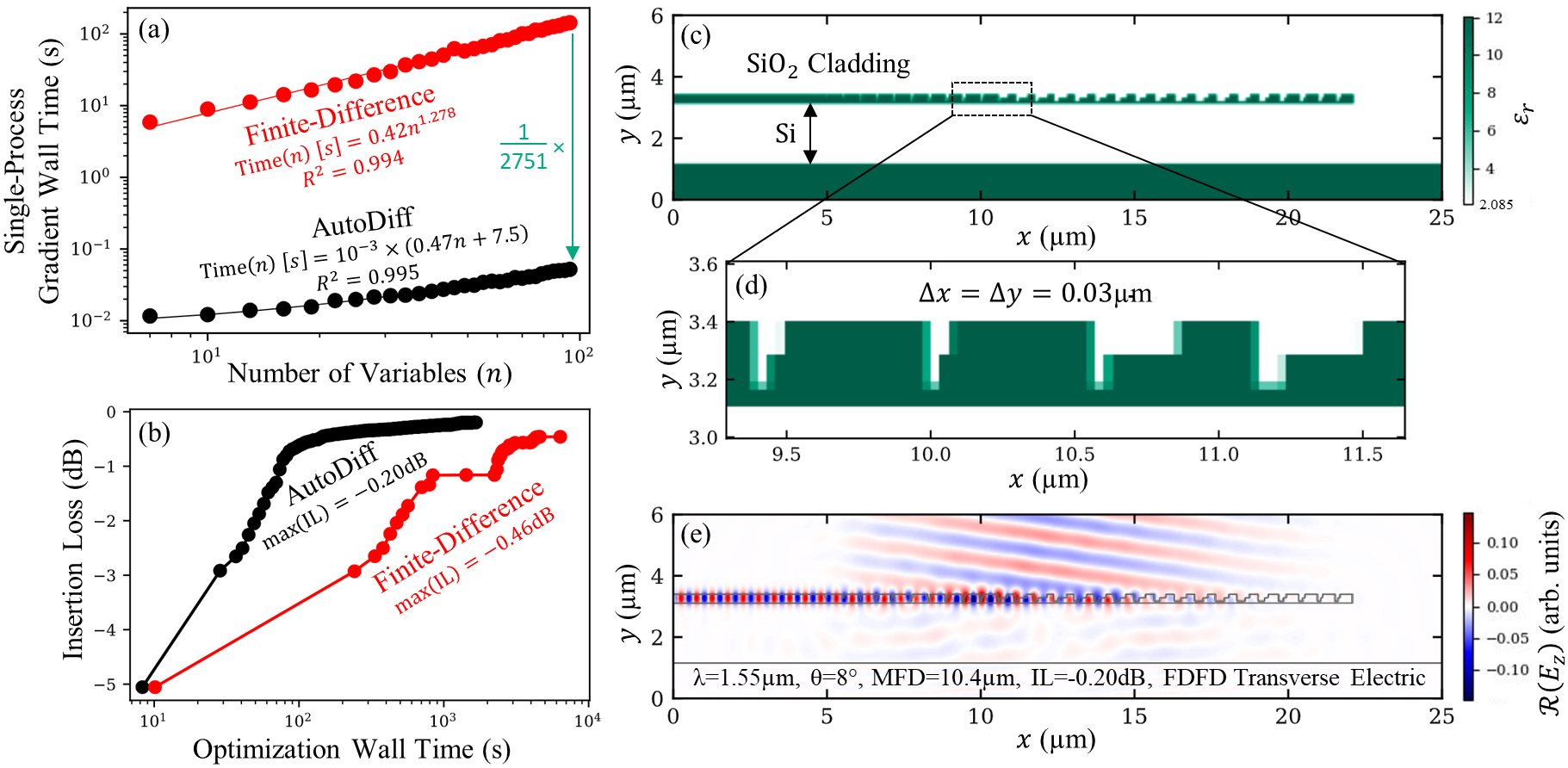}
    \caption{2-etch blazed grating coupler optimization using AutoDiff and Finite-Difference in the calculation of the adjoint method gradient. The gradient calculation time is accelerated by several orders-of-magnitude (a) resulting in greatly improved blazed grating performance and total optimization wall time (b) for a state-of-the-art blazed grating structure (c)-(e).}
    \label{fig:blazed_grating}
\end{figure}

The $N_g$ scattering elements of the blazed grating coupler are represented here via $2N_g$ two-dimensional rectangles. The permittivity layout of a blazed grating coupler with $N_g=30$ is shown in Fig.\,\ref{fig:blazed_grating}(c)-(d). The period of each scattering element and the widths of each of the 2 rectangles defining the scattering element are shape optimization variables (we constrain the two rectangles to share a boundary). Additional variables are the 2 etch depths defining the blazed grating, the buried oxide (BOX) thickness, and the position of the first scattering element relative to the simulation origin. This results in $n=3N_g+4$ total variables. 

For AutoDiff-enhanced calculation of the adjoint method gradient via Prop.\,1 (see Table\,\ref{tab:prop1and2}), the structure of the blazed grating in 2D FDFD transverse electric simulation is defined as combinations of AutoDiffGeo function objects (see Table\,\ref{tab:autodiffgeo}) using the following steps: (i) define the substrate using Rect2D, (ii) define the waveguide using Rect2D, (iii) define the scattering elements as pairs of Rect1D objects with variable width and period, (iv) extend the pairs of Rect1D objects to their respective etch depths using another pair of Rect1D broadcasting operations, (v) combine all scatterers using the differentiable union, (vi) subtract the unified scattering object from the waveguide object, (iv) scale all values of shape objects to the desired material relative permittivity of Si and SiO$_2$ ($\varepsilon_{\textrm{SiO}_2}=1.444^2$ and $\varepsilon_\textrm{Si}=3.4757^2$ were chosen, respectively). Note that we use the piecewise linear function described in Sec.\,\ref{sec:subpixelapprox}-\ref{sec:exp_prop1} as the underlying nonlinear function $\sigma_k$, meaning that the AutoDiffGeo representation of this structure, which consists of combinations of simple rectangles, \textbf{exactly} approximates volumetric boundary smoothing to within numerical precision. Therefore, we may invoke automatic differentiation in the optimization of this device without the need to convert the structure to a conventional volumetrically-averaged shape representation for validation. The optimization figure-of-merit is the coupling efficiency to an index-matched Gaussian optical fiber mode with MFD=10.4\textmu m, wavelength=1.55\textmu m, and tilt angle of $\theta=8^\circ$.

We perform 2 tests directly comparing AutoDiff to FD: (1) the time required to compute the adjoint method gradient versus the number of optimization variables $n$, and (2) the full optimization of the blazed grating coupler with $N_g=30$.

The results of (1) are shown in Fig.\,\ref{fig:blazed_grating}(a). We find that the AutoDiff calculation of the gradient is accelerated by several orders of magnitude relative to FD, up to as much as $2751$x for $n=94$ variables ($N_g=30$). Furthermore, we found that the FD data was better fit by a super-linear time scaling of $\textrm{Time}(n)\approx n^{1.278}$ for this problem, potentially limited by memory bandwidth. Meanwhile, the AutoDiff results exhibited the expected linear time scaling with $n$, and in fact is dominated by overhead up to about $n=16$ variables. 

There are several items of note: (i) $n=3N_g+4$, so we are able to scale the number of variables simply by varying the number of scattering elements in the simulation $N_g$. (ii) Here, we are only reporting the time to compute the adjoint method gradient (Eq.\,\ref{eq:1}, averaged over 10 attempts for each $n$ indicated in Fig.\,\ref{fig:blazed_grating}(a)), this does not include the time to perform the forward and adjoint simulations. (iii) The gradient calculation is performed on only a single CPU process for the most direct comparison of the two methods.\footnote{Note that the FD implementation currently scales better than the AutoDiff implementation with the number of MPI processes, meaning that the acceleration of AutoDiff reported here will be reduced by roughly a factor of $n_\textrm{processes}$ in more practical implementations invoking MPI. %Furthermore, as $n_\textrm{processes}$ is increased, we see the expected linear scaling with $n$, probably circumventing memory bandwidth issues. %Nevertheless, we anecdotally find that this scaling is not typically linear due to overhead in MPI communication. 
Furthermore, while not implemented here and overkill for this problem size, it is possible to use MPI with AutoDiff for improved scaling in a similar manner. Not to mention, AutoDiff is compatible with GPU acceleration by virtue of using PyTorch as the underlying AutoDiff library; here we only use CPU for a fair comparison.} 
(iv) The overall time to compute the gradient of both methods can be improved by limiting the material Jacobian calculation to design regions where the permittivity is locally perturbed by individual design variables. We did not consider any such local mappings because more overhead is required on part of the designer to define them, and because we are interested in generalizing our results to the worst-case situation (in terms of speed) where design variables perturb global regions of the simulation grid.

The superior acceleration of AutoDiff makes the gradient calculation approximately negligible in comparison to the forward and adjoint simulation times. This results in a significant improvement in total optimization wall time relative to Finite-Difference, which is shown in Fig.\,\ref{fig:blazed_grating}(b). Both optimizations use $n_\text{processes}=4$ MPI processes, $n=94$ variables ($N_g=30$), and the L-BFGS-B optimization routine available in SciPy. We allow both optimizations to proceed until convergence due to a $10^{-12}$ tolerance setting. Interestingly, the AutoDiff optimization is both faster and better behaved than FD, since the AutoDiff calculation of the gradient is numerically exact while the FD calculation is only an approximation limited by the chosen step size. The FD curve shown in Fig.\,\ref{fig:blazed_grating}(b) was the best optimization result from multiple swept finite-difference step sizes. As a consequence of improved numerical accuracy, the AutoDiff optimization proceeded for 365 L-BFGS-B iterations, and achieved a state-of-the-art insertion loss of just 0.20dB. Meanwhile, the FD optimization failed from loss of gradient accuracy after just 44 L-BFGS-B iterations, resulting in 0.46dB insertion loss. Furthermore, the AutoDiff optimization achieved these results in just 23\% of the total wall time of the FD case (a total acceleration of $4.3$x). The transverse electric field profile and final structure as defined on the simulation grid for the AutoDiff optimization are shown in Fig.\,\ref{fig:blazed_grating}(c)-(e).

Note that the wall time acceleration claimed here is primarily achieved because the FD version of the optimization is limited by the gradient calculation. % -- given an implementation where perturbations to the material Jacobian in the finite-difference case are tracked more carefully, the acceleration results can be far less monumental, albeit requiring more overhead on part of a designer. 
In cases where the material Jacobian FD calculation is negligible with respect to the forward/adjoint field solutions, the wall time acceleration would be less significant.\footnote{An implementation where non-zero entries of the material Jacobian along shape boundaries are tracked in a more principled manner would partially improve the speed of the finite-difference case, albeit requiring more overhead on part of the designer. We discuss this possibility in some more detail in our future work below in Sec.\,\ref{sec:conclusion}.} Nevertheless, the boost to final insertion loss achieved by AutoDiff due to the exact numerical calculation of the gradient represents a fundamental improvement offered by this method.

\subsection{Non-adiabatic transition taper in 2D}
\label{sec:taper2D}

Next, we consider a non-adiabatic waveguide taper, similar to the optimization originally proposed in Ref.\,\cite{michaels_leveraging_2018}. However, in this case we consider an input waveguide width of $w_\textrm{in}=0.5\textrm{\textmu m}$, an output waveguide width of $w_\textrm{out}=10.5\textrm{\textmu m}$, a total taper length of $L=23\textrm{\textmu m}$, and a wavelength of $\lambda=1.31\textrm{\textmu m}$. Furthermore, the taper has a unique geometrical parameterization in comparison to Ref.\,\cite{michaels_leveraging_2018}, described below. The taper is implemented in a 2D FDFD transverse magnetic simulation with silicon effective refractive index of $n_\textrm{Si,eff}=3.167$, as validated by an eigenmode simulation of a photonic SOI platform with silicon height of 300nm, full etch to define the waveguide, and silicon dioxide cladding. The figure-of-merit of the optimization is the mode-match efficiency of the fundamental mode injected at the narrow input to the fundamental mode at the wide output. A simple trapezoidal taper of this short length has insertion loss exceeding 4dB, therefore we invoke inverse design optimization of the shape's boundary to improve the coupling. We parameterize one side of the boundary (with mirror symmetry) using a linear function perturbed by a sine series with $n=100$ coefficients:
\begin{align}
    \label{eq:taper_boundary}f(x, v_1,...,v_{100}) &= \frac{w_\textrm{in}}{2} + \frac{(w_\textrm{out}-w_\textrm{in})}{2}\frac{x-x_0}{L}+\textrm{envelope}(x)\sum_{i=1}^{100} v_i\textrm{sin}\left(i\frac{\pi}{L}(x-x_0)\right)\\
    \textrm{envelope}(x) &= 0.1 + 0.45\cdot\left(1-\textrm{cos}\left(\frac{2\pi}{L}(x-x_0)\right)\right).
\end{align}
These equations are valid in the coordinate range $x\in[x_0,x_0+L]$ where $x_0=1\textrm{\textmu m}$ defines the input waveguide length. The full permittivity function in this range is then obtained via:
\begin{align}
    \varepsilon_\textrm{taper}(x,y,v_1,...,v_{100})=\varepsilon_{\textrm{SiO}_2}+(\varepsilon_\textrm{Si}-\varepsilon_{\textrm{SiO}_2})\sigma_k(f(x,v_1,...,v_{100})-y),
    \label{eq:taper_eps}
\end{align}
where $\sigma_k$ is the nonlinear function, here chosen to be the piecewise linear function with $k=1/\Delta x$ ($\Delta x=0.04\textrm{\textmu m}$).\footnote{We found that the choice of piecewise linear $\sigma_k$ for the taper boundary provided a slight edge in our optimization results, because other choices of $\sigma_k$ resulted in a discontinuity in the smoothened permittivity near the sensitive waveguide input region (which also used piecewise linear $\sigma_k$).} %because of its low volumetric approximation error with respect to many orientations of shape and grid boundary, shown in Fig.\,\ref{fig:quadrature_oneshot}.
In the rest of the simulation domain with coordinate range $x\in[0,x_0]\cup[x_0+L,25\textrm{\textmu m}]$, we use Rect2D objects with piecewise linear $\sigma_k$ to define the input and output waveguides (providing exact sub-pixel smoothing for waveguide mode injection and mode-matching). We note that Eq.\,\ref{eq:taper_eps} may be regarded as a general method to build differentiable shapes with nontrivial boundaries (after appropriate adjustments to function $f$ for a desired boundary parameterization), and can be further supplemented with differentiable union and intersection over additional shape objects.

\begin{figure}
    \centering
    \includegraphics[width=0.9\textwidth]{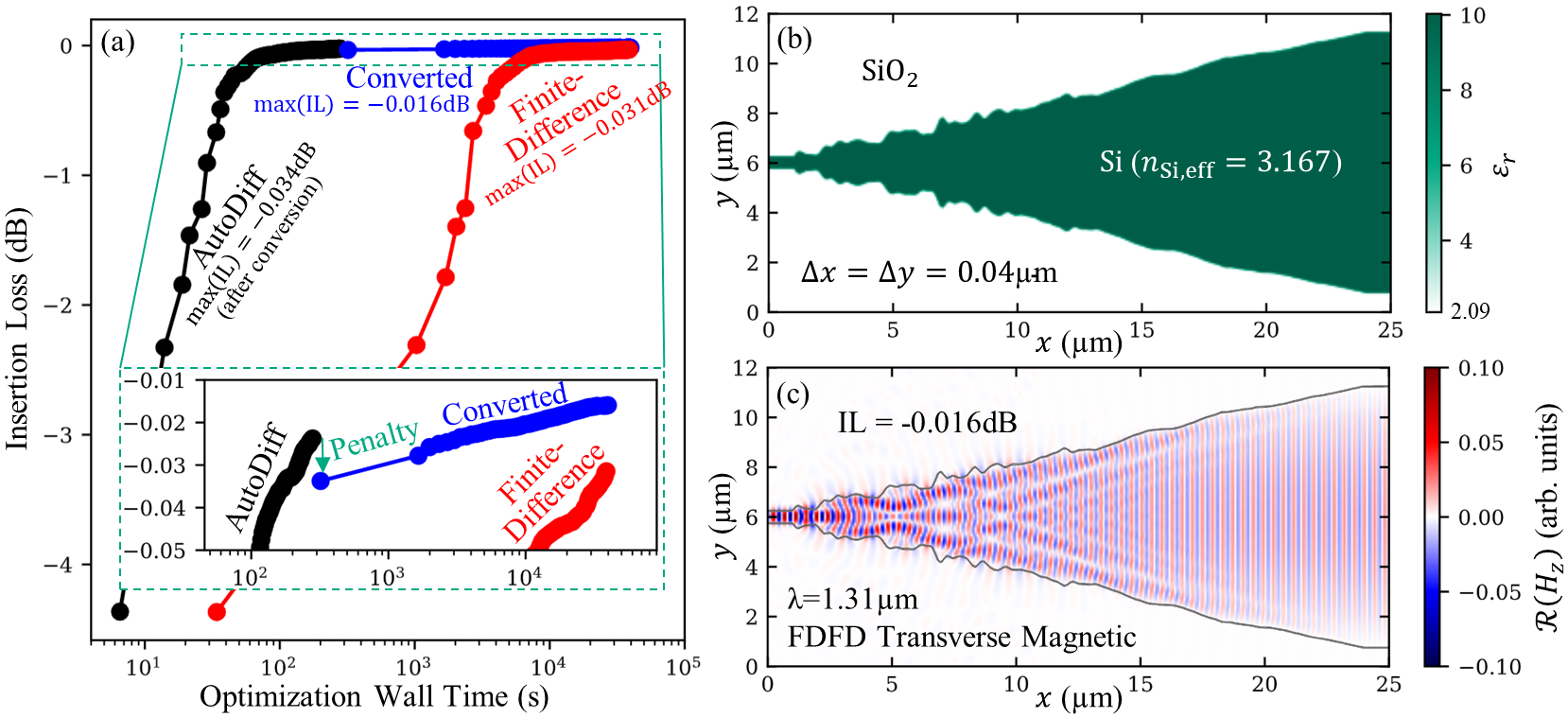}
    \caption{2D transition taper optimization comparing the performance of Finite-Difference and AutoDiff methods to compute the adjoint method gradient. We allow both methods to proceed for approximately the same wall time (a), and present the final optimized result of the AutoDiff/Converted case (b)-(c).}
    \label{fig:taper2D}
\end{figure}

We perform an AutoDiff-enhanced optimization of the taper, in comparison to standard finite-difference in Fig.\,\ref{fig:taper2D}(a). In the FD optimization, we represent the taper shape as a polygon with 200 vertices (per side) with coordinates sampled uniformly from Eq.\,\ref{eq:taper_boundary} above. Both optimizations use $n_\text{processes}=16$ MPI processes, $n=100$ variables, and the L-BFGS-B optimization routine available in SciPy with default settings. Since the AutoDiffGeo shape representation admits only an approximation of the volumetrically-averaged sub-pixel smoothing strategy, the structure suggested by the optimization must be converted to the more precise representation, incurring a penalty to the figure-of-merit after the change. Therefore, as suggested by Prop.\,1 in Sec.\,\ref{sec:subpixelsmoothing}, we perform a FD refinement optimization on the converted structure. In particular, we allow the standard FD optimization to proceed for 100 L-BFGS-B iterations. We allow the AutoDiff optimization to proceed for 100 L-BFGS-B iterations, convert it, then allow it to run for 100 L-BFGS-B iterations using the finite-difference scheme. The AutoDiff/converted optimization runs for about the same wall time as the full FD optimization because the FD optimization's wall time is highly limited by the gradient calculation. This is because each variable $v_j$ in the sine series parameterization in Eq.\,\ref{eq:taper_boundary} affects large regions of the simulation domain, and therefore nearly the entire simulation grid needs to be updated in $n=100$ finite-difference perturbations of the material Jacobian. By contrast, the AutoDiff optimization is only limited by simulation time, ultimately making iterations approximately 10x faster. 

We find that both optimizations reached nearly 0 insertion loss at the simulation wavelength of interest. However, for a given optimization wall-time, the performance of the AutoDiff/converted optimization is always better than the standard FD optimization. Moreover, the penalty incurred by conversion of the AutoDiff optimized structure to the exact volumetrically-averaged version is negligible by experimental standards ($\sim0.02\textrm{dB}$ conversion penalty), indicating that the physically-guided choice of AutoDiffGeo nonlinear function introduces low error, at least for non-resonant structures of this type. If a designer is willing to allow for these minor penalties on the optimized design, large improvements to the total optimization wall time can be achieved.

As a final point, we note that there is an alternative parameterization of this problem where the FD optimization wall time can be improved by directly parameterizing the coordinates defining the vertices of the taper boundary, and manually computing dielectric perturbations only within the vicinity of those vertices, as was utilized in Ref.\,\cite{michaels_leveraging_2018}. We chose the Fourier-decomposed parameterization here, given in Eq.\,\ref{eq:taper_boundary}, for 2 reasons: (i) it provides an example where design variables map to global changes in the material distribution, and (ii) the AutoDiffGeo shape definitions are currently incompatible with directly defining concave polygons. Global/Fourier-encoded feature maps, as in (i), are commonly used in photonic shape optimization \cite{michaels_inverse_2018, michaels_hierarchical_2020, hooten_adjoint_2020, sun_adjoint_2023} because they represent physically-guided and intuitive parameterizations, and natively maintain fabrication compatibility by cutting off high-frequency features. As for (ii), AutoDiff-compatible concave polygon parameterizations can be implemented with a conformal mapping \cite{driscoll_schwarz-christoffel_2002}, which represents a future work direction. Alternatively, one may leverage union/intersection operations (Eq.\,\ref{eq:union}--\ref{eq:intersection}) for concave polygons. Nevertheless, the parameterization chosen here allows us to directly compare the two methods on equal grounds.

\subsection{Polarization-splitting grating coupler in 3D}
\label{sec:PSGC}

Finally, we consider the polarization-splitting grating coupler (PSGC), which we have  detailed and optimized in a previous study \cite{sun_adjoint_2023}. The PSGC permittivity layout before optimization was discussed in Sec.\,\ref{sec:prop2} and visualized in Fig.\,\ref{fig:psgc_full_jac_smh}(a). The simulation domain of the PSGC has dimensions of $17.5\textrm{\textmu m} \times 17.1\textrm{\textmu m} \times 2.6\textrm{\textmu m}$ and consists of $\sim$29 million FDTD grid voxels at $\Delta x=30\textrm{nm}$ uniform mesh size. The individual PSGC scatterers are generated using the parameterization discussed in Eq.\,\ref{eq:PSGCboundary}. There are several additional global parameters, e.g. describing aspect ratio of scatterers, described in Ref.\,\cite{sun_adjoint_2023}. Overall, there are 325 parameters.
The PSGC couples light between a single-mode fiber with mode-field diameter (MFD) of 9.2\textmu m and silicon ridge waveguides in the O-band.  The Figure-of-Merit (FOM) of the PSGC optimization problem is defined as weighted sum of 3 single-wavelength FOMs at 1305nm, 1310nm, and 1315nm respectively.  Each single-wavelength FOM is defined as the sum of two terms: the lower coupling efficiency of S- and P-polarization, minus the square of the difference of the S- and P-polarization coupling efficiencies:
\begin{equation}
\begin{split}
\textrm{FOM}=&c_{1305}[\min(S_{1305},P_{1305})-\zeta(S_{1305}-P_{1305})^2] \\
&+c_{1310}[\min(S_{1310},P_{1310})-\zeta(S_{1310}-P_{1310})^2] \\
&+c_{1315}[\min(S_{1315},P_{1315})-\zeta(S_{1315}-P_{1315})^2]\;,
\end{split}
\end{equation}
where $\zeta=1000$ and $(c_{1305},c_{1310},c_{1315})=(2,1,2)$. 

\begin{figure}
    \centering\includegraphics[width=0.5\linewidth]{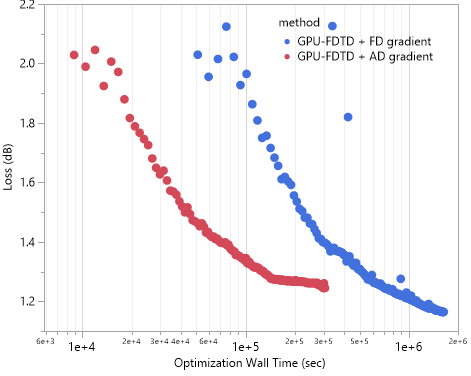}
    \caption{PSGC 1310 nm S-polarization loss vs. optimization wall time, optimized by the FD method and AutoDiff methods, respectively.}
    \label{fig:PSGC_loss_walltime_smh}
\end{figure}

We compare optimization results of the AutoDiff method with that of the baseline finite-difference (FD) method. For AutoDiff-enhanced optimization, we use Prop.\,2 from Table\,\ref{tab:prop1and2}, where the AutoDiff-compatible representation of the PSGC layout is only used in the adjoint method gradient calculation.  Fig.\,\ref{fig:PSGC_loss_walltime_smh} shows the optimized PSGC's S-pol 1310 nm loss versus the wall time using both FD and AutoDiff methods. Note that we use a GPU-accelerated FDTD solver for forward/adjoint simulations in both cases \cite{psun_emopt_2023, peng_gpuemopt_2023}. The FD method calculates the volume-weighted permittivity gradient with respect to all the design parameters with finite difference scheme, which means that the grid meshing needs to be repeated for a total of $325\times3=975$ times, where the factor 3 is to account for the half grid difference in effective permittivity along the X/Y/Z axes, respectively.  After the volume-weighted permittivity gradient is calculated, the FOM gradient can then be calculated by multiplying the permittivity gradient with the forward- and adjoint- FDTD field matrices and reducing the resulting matrices to scalars for each of the design parameters as in Eq.\,\ref{eq:1}.  Since only part of the silicon slab will change under the design parameter perturbation, we only need to re-mesh a small fraction ($\sim$ 10\%) of the simulation domain to save time.  On a powerful server with 16-socket $\times$ 18-core CPUs, it takes $\sim$ 2 sec to mesh the downsized domain, and $\sim$1 sec to multiply and reduce the permittivity gradient and the field matrices in X/Y/Z directions.  The total time to calculate all 325 elements of the FOM gradient is then: $325\times3\times2\textrm{ sec}+325\times1\textrm{ sec}=2275\textrm{ sec}=38\textrm{min}$. On the other hand, the AutoDiff method calculates the gradient via Eq.\,\ref{eq:AD}-\ref{eq:jacmu}, where the PSGC layout needs to be generated only one time with the AutoDiff-compatible representation. %the gradient of a scalar loss function, which we handily pick as the inner product of the full permittivity gradient and the forward- and reverse-FDTD field matrices.  
On a single 4-core CPU, it takes only $\sim$36 sec to calculate all 325 elements of the FOM gradient with the AutoDiff method, or an acceleration of 63$\times$. Thus, we observe a total wall time acceleration of $\sim4\times$ in Fig.\,\ref{fig:PSGC_loss_walltime_smh} because the AutoDiff optimization is limited only by forward and adjoint simulations, whereas the FD optimization is limited primarily by the gradient calculation. 

\begin{figure}
     \centering
     \includegraphics[width=1\linewidth]{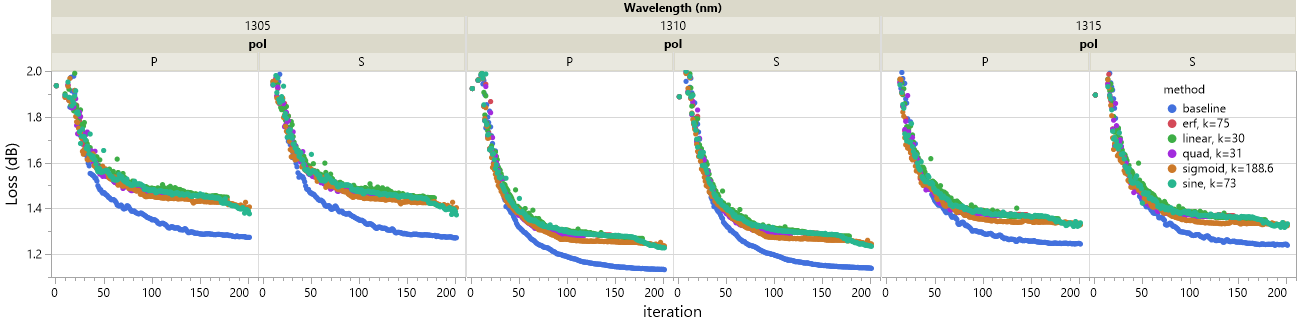}
     \caption{PSGC losses at 1305/1310/1315 nm and S/P polarizations, optimized by FD- and AD-methods, respectively.}
     \label{fig:PSGC_fullopt_FD_vs_AD_smh}
 \end{figure}

Fig.\,\ref{fig:PSGC_fullopt_FD_vs_AD_smh} overlays the optimized PSGC's S- and P-polarization losses at the 3 wavelengths using the FD method and the AutoDiff method.  For the AutoDiff method, we consider all 5 nonlinear functions from Table\,\ref{tab:nonlinearfunctions}, where the inverse-length-scale $k$ is set to their respective optimal values obtained from minimizing the RMSE in permittivity derivatives from Fig.\,\ref{fig:psgc_jac_fitting_smh}.  We observe that error in the AutoDiff method accumulates as the optimization goes on, but the impact on PSGC loss is limited to about 0.1 dB for all 5 nonlinear functions.  It is also worth noting that optimizations using the quad, erf, and linear functions terminate at fewer iterations than the maximum of 200 (quad at iteration 117, erf at iteration 145, linear at iteration 178) due to x-tolerance condition being satisfied, which could be a sign of lost accuracy in gradient. This is an expected consequence of using Prop.\,2 for AutoDiff optimizations, since this method uses different representations of device geometry in the simulations versus the gradient calculation, resulting in a noisy approximation of the FD gradient.

\section{Discussion and Conclusion}
\label{sec:conclusion}
In this work we demonstrated methods to accelerate shape optimization for photonic inverse design on rectilinear simulation grids by invoking automatic differentiation capability in instantiations of structural material distributions. To do so, we developed AutoDiffGeo, a library of highly extensible, closed-form differentiable mappings from structural shape parameters to elementary shape primitives, as well as differentiable logic operations that support multiple materials. Although, our AutoDiffGeo definitions currently do not perfectly satisfy sub-pixel smoothing on rectilinear simulation grids for arbitrary geometries, there are principled methods to imitate sub-pixel smoothing, allowing optimizations to provide physically meaningful solutions that introduce minimal additional discretization error to field solutions and objective calculations. Furthermore, we showed that common structural representations such as cuboids and planarity transformations achieve zero error with respect to volumetric averaging. We demonstrated the effectiveness AutoDiff-enhanced shape optimization on three integrated photonics applications, ultimately finding $>50\times$ acceleration in the gradient calculation and $>4\times$ acceleration in total optimization wall time. 

There are several avenues for future work. Firstly, we note that the acceleration offered by AutoDiff for shape optimization is primarily due to the use of closed-form expressions for $\varepsilon, \mu$. Doing so removes the core problem present in conventional shape optimization representations, where the material Jacobian is computed by manual finite-differences. An enticing alternative possibility to improve the finite-difference method is to find a more scalable, principled method to track non-zero entries in the material Jacobians of arbitrary combinations of shapes, which would greatly reduce the difficulty of computing Eq.\,\ref{eq:1} (at least, for cases where variables map to only local regions of the designable grid). Nevertheless, the differentiable parameterized structures presented in this provide exact numerical gradients for Eq.\,\ref{eq:1} as opposed to a finite-difference approximation, and furthermore, they are useful outside of the computation of Eq.\,\ref{eq:1}, e.g. for deep learning models that take into account structural geometry \cite{lu_physics-informed_2021, zhelyeznyakov_large_2023}. 

In this paper we compared the material distributions defined by AutoDiffGeo shape representations to a basic arithmetic averaging of materials over rectilinear pixels \cite{michaels_leveraging_2018}. To extend the AutoDiffGeo method to the more sophisticated, exact sub-pixel smoothing method \cite{ farjadpour_improving_2006, oskooi_accurate_2009, oskooi_meep_2010}, one needs to consider (i) an additional approximation of the harmonic average of material values per grid cell, and (ii) a mapping of arithmetic/harmonic averages to an anistropic material tensor that takes into account the orientation of the shape's boundary with grid cell. We believe that (i) should be achievable with alternative choices of nonlinear building block functions $\sigma_k$. As for (ii), we showed in Sec.\,\ref{sec:subpixelsmoothing} that even the simple arithmetic average has strong dependence on shape/grid cell orientation. Therefore, implementing (ii) will likely be accompanied by an extension to nonlinear functions $\sigma_k$ that take into account shape/grid cell orientations. It is not clear at the time of this work whether (i) and (ii) can be done in a way that is compatible with AutoDiff, and that avoids supersampling the grid cell (e.g., convolutional smoothing).

A final future work direction is in the implementation of arbitrary convex/concave polygons using AutoDiff-compatible transformations. Currently our Poly2D and Polar2D functions from Table\,\ref{tab:autodiffgeo} only support convex and star-convex polygons, respectively. Arbitrary concave polygons can be constructed through union / intersection operations, but this may lack convenience in some situations. A possible method to implement arbitrary concave polygons would be through a Schwarz-Christoffel conformal mapping \cite{driscoll_schwarz-christoffel_2002}, possibly at the expense of more difficult computation in the gradient calculation.

\printbibliography

\appendix

\section{AutoDiffGeo shape definitions}
\label{app:shape_defs}
Several AutoDiffGeo shape primitives function definitions are defined in Table\,\ref{tab:autodiffgeo}. Parameters listed in this table correspond to the examples in Sec.\,\ref{sec:diffgeo}, Fig.\,\ref{fig:shape_examples}. Efficient coded versions of these definitions leveraging convenient vectorization over arrays of input coordinates are implemented with the PyTorch API in our custom AutoDiffGeo module, integrated with EMopt \cite{hooten_emopt_2023}.

\begin{sidewaystable}
%\begin{table}
\centering
\begin{threeparttable}
    \caption{AutoDiffGeo definitions}
    \centering
    \def\arraystretch{1.5}
    \begin{tabular}{llll}
		\toprule
		Shape                                     & Equation                                                          & Notes & Example Parameters (Fig.\,\ref{fig:shape_examples})  \\
		\midrule
		$\textrm{Step1D}(x,x_0)$                  & $\sigma_k(x-x_0)$                                                 & $^{1,2}$ & $x_0=-0.4$ \\
		$\textrm{Rect1D}(x,x_0,x_1)$              & $\sigma_k(x-x_0) \cdot \sigma_k(x_1-x)$                           & $^{1,2}$ & $x_0=-0.4, x_1=0.5$ \\
		$\textrm{Rect2D}(x,y,x_0,y_0,x_1,y_1)$    & $\textrm{Rect1D}(x,x_0,x_1)\cdot \textrm{Rect1D}(y,y_0,y_1)$      & $^{1,2}$ & $x_0=-0.4, x_1=0.5, y_0=-0.4, y_1=0.7$ \\
		$\textrm{RectdD}(r^1,...,r^d, r^1_0,...,r^d_0, r^1_1, ..., r^d_1)$ & $\textrm{Rect1D}(r^1,r^1_0,r^1_1)\cdot ... \cdot \textrm{Rect1D}(r^d,r^d_0,r^d_1)$ & $^{1,2}$ & -- \\
		$\textrm{Step2D}(x,y,\vec{n},x_0,y_0)$    & $\sigma_k(n_x(x-x_0)+n_y(y-y_0))$                                 & $^{1,2,3}$ & $x_0=0.2, y_0=0.2$, $\vec{n}=[1/2, \sqrt{3}/2]$ \\
		$\textrm{Poly2D}(x,y,x_0^1,...,x_0^N,y_0^1,...,y_0^N)$       & $\prod_{i=1}^{N} \textrm{Step2D}(x,y,\vec{n}_{i\rightarrow i+1}, x_0^i, y_0^i)$ & $^{1,2,4}$ &  $(x_0^1,x_0^2,x_0^3,y_0^1,y_0^2,y_0^3)=(-0.7,0.7,0,0.6, 0.5, -0.5)$ \\
		$\textrm{Circ2D}(x,y,R,x_0,y_0)$          &  $\sigma_k\left(R-r(x,y,x_0,y_0)\right)$                           & $^{1,2,5}$             &  $R=0.5$, $x_0=0.0$, $y_0=-0.5$\\
		$\textrm{Polar2D}(x,y,R,\delta,x_0,y_0,\alpha)$ & $\sigma_k\left(R(1+\delta \cos(\alpha\theta(x,y,x_0,y_0)))-r(x,y,x_0,y_0)\right)$                & $^{1,2,5,6}$ & $R=0.5$, $\delta=0.2$, $x_0=0.0$, $y_0=0.0$, $\alpha=4$  \\
        \midrule
        $\textrm{GeneralCartesian2D}(x,y,v_1,...,v_n|f)$ & $\sigma_k(\pm(f(x,v_1,...,v_n)-y))$                        & $^{1,2}$ & -- \\
        $\textrm{GeneralPolar2D}(\theta, r, v_1,...,v_n|f)$ & $\sigma_k(\pm(f(\theta, v_1,...,v_n)-r))$                        & $^{1}$ & -- \\
		\bottomrule
    \end{tabular}
    \begin{tablenotes}
        \item $^1$All definitions make use of a nonlinear function $\sigma_k(\cdot):\mathbb{R}\rightarrow[0,1]$. Examples are provided in Appendix\,\ref{app:nonlinearities}, Table\,\ref{tab:nonlinearfunctions}.
        \item $^2$All functions here are defined with respect to scalar input Cartesian coordinates $x,y$ (or $r^1,...,r^d$ for $d$-dimensional versions). In practice, we may pass arrays of coordinate values, and make use of packages like NumPy or PyTorch for convenient vectorization over the AutoDiffGeo shape definitions. Please see \cite{hooten_emopt_2023} for efficient coded definitions when passing arrays of coordinates.
        \item $^3$ $\vec{n}$ is assumed to be a 2-element unit-normal that defines the direction of Step2D transition. $\vec{n}=[n_x, n_y]$ such that $n_x$ and $n_y$ are the elements in the $x$ and $y$ Cartesian directions.
        \item $^4$ $\vec{n}_{i\rightarrow i+1}$ is a function that takes the unit-normal between coordinates defining vertices of the polygon (i.e., the normal of $(x_0^i,y_0^i)\rightarrow(x_0^{i+1},y_0^{i+1})$). The edge case is given by $\vec{n}_{N\rightarrow N+1}=\vec{n}_{N\rightarrow 1}$.
        \item $^5$ $r(x,y,x_0,y_0)=\sqrt{(x-x_0)^2+(y-y_0)^2}$.
        \item $^6$ $\theta(x,y,x_0,y_0)=\tan^{-1}\dfrac{y-y_0}{x-x_0}$.
    \end{tablenotes}
    \label{tab:autodiffgeo}
\end{threeparttable}
%\end{table}
\end{sidewaystable}

\section{Extended discussion of union and intersection}
\label{app:union}

\subsection{Non-binary material systems}
\label{app:union_nonbinary}
In the main text we only considered shape optimization with respect to binary material systems. In general, our methods may be directly extended to non-overlapping non-binary material systems simply by defining shapes with different coefficients representing their material values. However, we run into trouble when shapes with multiple levels of material values overlap. In particular, the union and intersection operations from Eq.\,\ref{eq:union}-\ref{eq:intersection} in the main text can only be applied to sets of shapes with two levels of material values. In order to consider overlapping non-binary materials, union and intersection must be applied iteratively. One such way is described in this section.

Suppose we are provided $K$ sets of shape primitives, where each set consists of $N_k$ members defined on a size $M$ coordinate grid. For example, the $k$-th set can be denoted $\{s_1^k,...,s_{N_k}^k\}$ where $s_i^k\in[0,1]^M$ for any given $i\in\{1,...,N_k\}$ represents the bounded continuous shape primitive elements. Suppose we associate a peak material value with each of the $K$ shape sets: e.g., the $k$-th set has material level $\varepsilon^k$, relative to background. Without loss of generality, we assume that the material values are sorted: $\varepsilon^{k-1}\leq \varepsilon^{k}$ for all $k$. The algorithm to perform union over all $K$ sets proceeds as follows in Alg.\,\ref{alg:cap}.
\begin{algorithm}[ht]
\caption{An algorithm to perform union over non-binary material structures with $K+1$ total levels, where $K>1$.}\label{alg:cap}
\begin{algorithmic}
\State $s\gets \textrm{union}(s_1^1,...,s_{N_1}^1)$
\For{$k=2, ..., K$} 
    \State $s^k\gets\textrm{union}(s_1^k,...,s_{N_k}^k)$
    \State $s\gets \textrm{union}\left(\dfrac{\varepsilon^{k-1}}{\varepsilon^{k}}s, s^k\right)$
\EndFor
\State $s\gets \varepsilon^Ks$
\end{algorithmic}
\end{algorithm}

\begin{figure}
    \centering
    \includegraphics[width=0.9\textwidth]{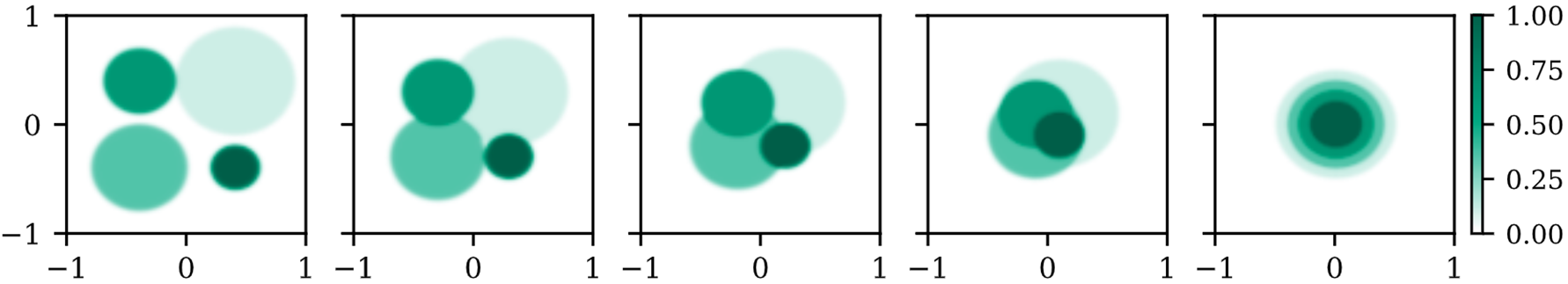}
    \caption{Union applied to sets of shapes consisting of more than one material level. From left to right, circles with different material values are gradually brought from their respective quadrants to the center.}
    \label{fig:nonbinary_union}
\end{figure}

Note that this algorithm will give priority to the sets of shapes with largest material value. An example is shown in Fig.\,\ref{fig:nonbinary_union}, where from the left-most panel to the right-most panel, we overlap circles from each of the 4 quadrants with respective material levels $0.1, 0.34, 0.62, 1.0$ and background material value of 0. We emphasize that this algorithm maintains automatic differentiability, and therefore benefits from the accelerated shape optimization proposed in this work.

%Then for  $\{s_1,...,s_N\}$ where $\forall i$, $s_i\in[0,1]^M$. However, through some additional bookkeeping we can apply the same operations to non-binary material systems. Suppose we are given two sets of shapes 
\subsection{Union and intersection ``truth tables'' on continuous scale}
\label{app:union_truthtable}

In this appendix we provide a more detailed explanation for the choice of differentiable union and intersection functions in Eq.\,\ref{eq:union} and Eq.\,\ref{eq:intersection}, respectively. We will first state several options, then consider an instructive use case. Let $s_1,...,s_N$ represent $N$ shape primitives such that for any $i=1,...,N$, $s_i\in[0,1]^M$ where $M$ is the size of the coordinate grid where each shape was sampled. We define 3 differentiable union and intersection function options below such that for any $j=1,2,3$, $\textrm{union}_j:[0,1]^{NM}\rightarrow[0,1]^M$ and $\textrm{intersection}_j: [0,1]^{NM}\rightarrow[0,1]^M$:
\begin{align}
    \textrm{union}_1(s_1, ..., s_N) &= \sigma_k\left(\sum_{i=1}^N s_i - \frac{1}{2}\right)\label{eq:union2} \\
    \textrm{union}_2(s_1, ..., s_N) &= s_N + (1-s_N)\textrm{union}_2(s_1,...,s_{N-1}), \quad\textrm{union}_2(s_1)=s_1\label{eq:union3}\\
    \textrm{union}_3(s_1, ..., s_N) &= \min\left\{1,\,\sum_{i=1}^N s_i\right\} \label{eq:union1}
\end{align}
\begin{align}
    \textrm{intersection}_1(s_1, ..., s_N) &= \sigma_k\left(\sum_{i=1}^N s_i - (N-\frac{1}{2})\right) \label{eq:intersection2} \\
    \textrm{intersection}_2(s_1, ..., s_N) &= \prod_{i=1}^N s_i \label{eq:intersection3}\\
    \textrm{intersection}_3(s_1, ..., s_N) &= \max\left\{N-1,\,\sum_{i=1}^N s_i\right\} - (N-1) \label{eq:intersection1}
\end{align}

where in Eq.\,\ref{eq:union2} and Eq.\,\ref{eq:intersection2} we made use of a [0,1]-bounded smooth nonlinear function $\sigma_k$; we note that Eq.\,\ref{eq:union3} is defined recursively; and in Eq.\,\ref{eq:union1} and Eq.\,\ref{eq:intersection1} we have repeated the functions originally given in Eq.\,\ref{eq:union} and Eq.\,\ref{eq:intersection};.

\begin{figure}
    \centering
    \includegraphics[width=0.9\textwidth]{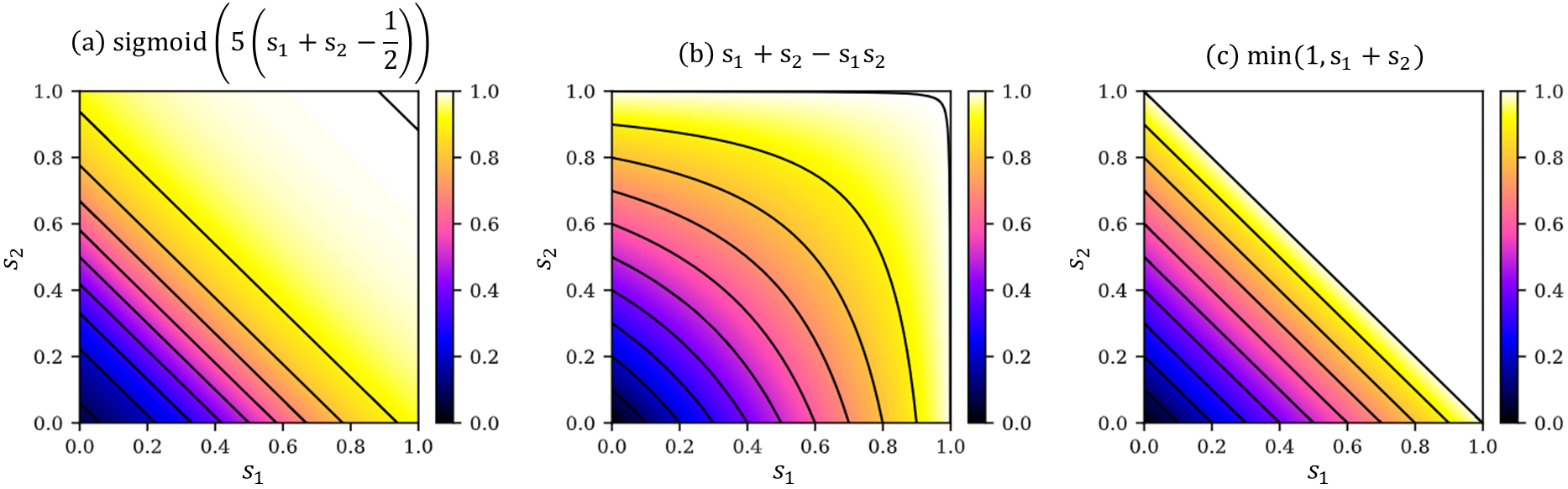}
    \caption{Options for differentiable union between two shapes with corresponding continuous material values $s_1$ and $s_2$ at any given simulation pixel. Each option can be thought of as a version of a logical truth table on continuous scale.}
    \label{fig:truthtable}
\end{figure}

Without loss of generality and for purposes of instruction, we will consider the union equations given in Eq.\,\ref{eq:union2}-\ref{eq:union1} applied to 2 smooth shape primitives $s_1,s_2\in[0,1]^M$.  Since each function is applied element-wise to each of the $M$ coordinate values of $s_1,s_2$, we can generally consider the output of each function when $M=1$. All possible inputs and outputs of this situation are visualized in the color plots of Fig.\,\ref{fig:truthtable} where 10 equally spaced contours  in the bounds [0,0.999] are plotted as guides for the eye. We may now directly compare the values in Eq.\,\ref{fig:truthtable} to the logical union (logical ``OR'') applied to 2 binary bit values, $b_1,b_2$ in Table.\,\ref{tab:truthtable}. We find that each function in Fig.\,\ref{fig:truthtable}(a)-(c) generally meets the criteria of Table\,\ref{tab:truthtable} at the bounds of $s_1,s_2\in\{0,1\}$ with analogy to $b_1,b_2$, respectively, except for the case of Fig.\,\ref{fig:truthtable}(a) which does not output the desired value of 1 for the cases where $s_1=1, s_2=0$ and $s_1=0, s_2=1$, but we can get arbitrarily close to the desired output by increasing the value of $k=5$ to something larger. The question now becomes, what output do we desire in the continuous region for $s_1,s_2\in(0,1)$? This is somewhat a subjective matter, as each function in Eq.\,\ref{fig:truthtable} has its pros and cons. For example, the value associated with coordinates along the boundaries of the AutoDiffGeo shape primitives is 0.5. Therefore, when the boundaries of 2 shapes exactly overlap such that $s_1,s_2=0.5$, then the desired output of $\textrm{union}(s_1,s_2)=1$ in the sense that there is no longer a boundary at that location (though this assumes that the interiors of the shapes are in opposite directions of the shared boundary). Only Fig.\,\ref{fig:truthtable}(c) fulfills this criterion exactly, though Fig.\,\ref{fig:truthtable}(a) can come arbitrarily close with different choice of $\sigma_k$ or large $k$ (amounting to a hyperparameter choice). Fig.\,\ref{fig:truthtable}(b) requires the shapes to fully overlap before the union takes on value of 1. However, Fig.\,\ref{fig:truthtable}(a) and Fig.\,\ref{fig:truthtable}(b) have the desirable quality that they are continuously differentiable across the entire domain of $s_1,s_2\in[0,1]$, while Fig.\,\ref{fig:truthtable}(c) is only piecewise differentiable. As mentioned in the main text, this is still compatible with AutoDiff, and generally the discontinuities in shape Jacobia will wash out after reduction in the full calculation of the gradient, but still may be worth consideration in contexts not treated here.

As a final note, while the intersection functions in Eq.\,\ref{eq:intersection2}-Eq.\,\ref{eq:intersection3} were not discussed here, the conclusions reached by our treatment in Fig.\,\ref{fig:truthtable} are similar.

\begin{table}
	\caption{Logical union truth table}
	\centering
	\begin{tabular}{ccc}
		\toprule
		  $b_1$ & $b_2$ & $b_1$ OR $b_2$ \\
		\midrule
            1 & 1 & 1 \\
            1 & 0 & 1 \\
            0 & 1 & 1 \\
            0 & 0 & 0 \\
		\bottomrule
	\end{tabular}
	\label{tab:truthtable}
\end{table}

\section{Nonlinear function definitions}
\label{app:nonlinearities}
\begin{figure}
    \centering
    \includegraphics[width=0.75\textwidth]{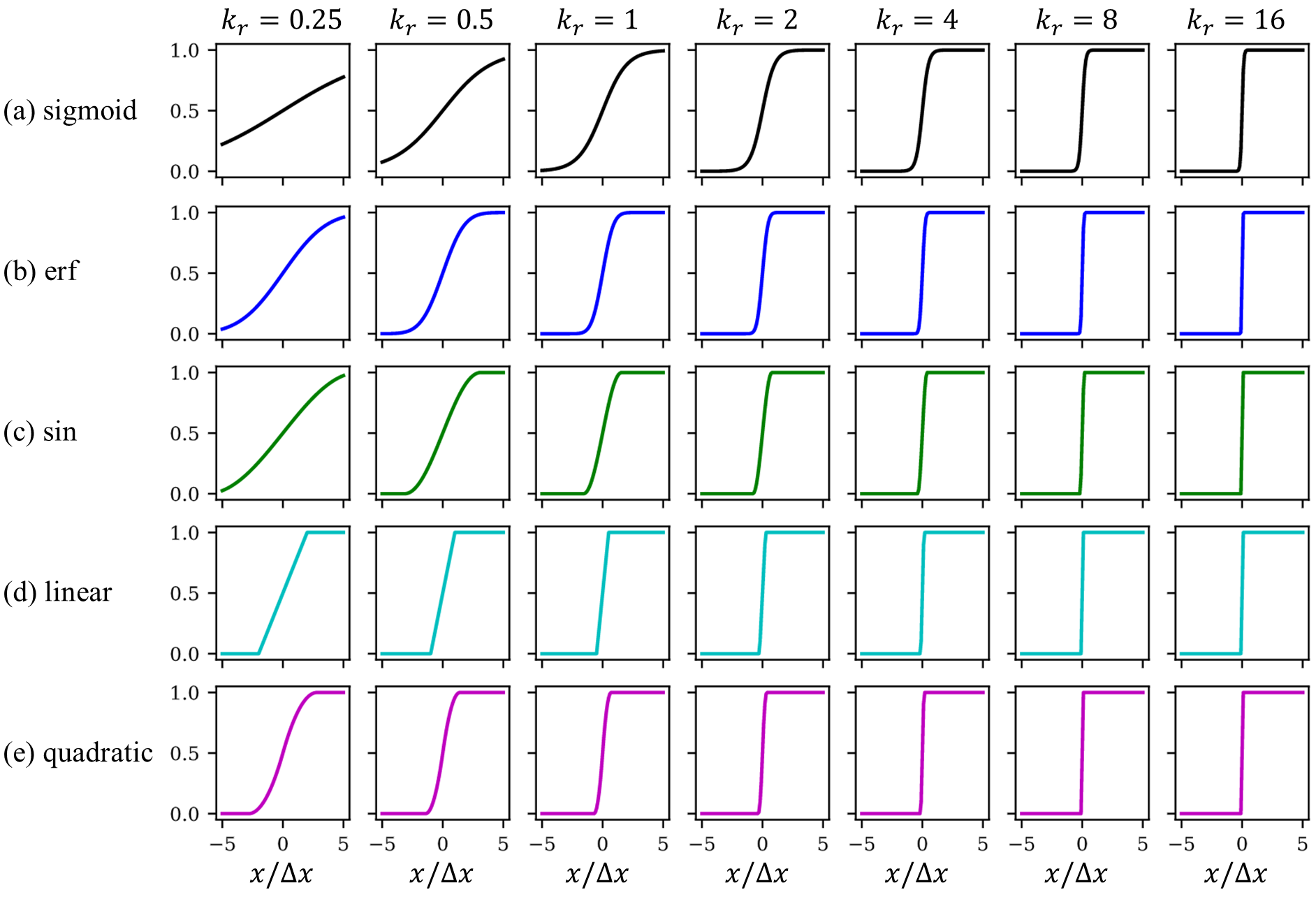}
    \caption{Several choices of $\sigma_k$ with characteristic inverse-length-scale $k=k_r/\Delta x$. Function definitions provided in Table\,\ref{tab:nonlinearfunctions}}
    \label{fig:nonlinearities}
\end{figure}

Various (piecewise-)smooth, [0,1]-bounded function definitions that were used in the main body of the paper, $\sigma_k(\cdot)$, are provided in Table.\,\ref{tab:nonlinearfunctions} and plotted in Fig.\,\ref{fig:nonlinearities}.

\begin{table}
	\caption{Nonlinear function $\sigma_k$ definitions}
	\centering
        \def\arraystretch{2}
	\begin{tabular}{ll}
		\toprule
		  $\sigma_k(x)$    & Function Definition  \\
		\midrule
		$\textrm{sigmoid}(k,x)$           & $(1+\exp(-kx))^{-1}$ \\
		$\textrm{erf}(k,x)$              & $\dfrac{1}{2}\left(1+\textrm{Erf}(kx)\right)$ \\ 
            $\textrm{sin}(k,x)$               & $\begin{cases}
                                        0        & \quad\quad kx < -\pi/2 \\
                                        \frac{1}{2}(1+\sin(kx))  & -\pi/2 \leq kx \leq \pi/2 \\
                                        1        & \quad\quad kx > \pi/2
                                      \end{cases}$ \\
            $\textrm{linear}(k,x)$        & $\max\{0, \min\{1, kx+\frac{1}{2}\}\}$ \\
            $\textrm{quadratic}(k,x)$         & $\begin{cases}
                                        0                               & kx < -\frac{1}{\sqrt{2}} \\
                                        \left(\frac{1}{\sqrt{2}}+kx\right)^2 & -\frac{1}{\sqrt{2}} \leq kx < 0 \\
                                        1-\left(\frac{1}{\sqrt{2}}-kx\right)^2  & 0 \leq kx \leq \frac{1}{\sqrt{2}} \\
                                        1        & kx > \frac{1}{\sqrt{2}} \\
                                      \end{cases}$ \\
		\bottomrule
	\end{tabular}
	\label{tab:nonlinearfunctions}
\end{table}

\end{document}